\documentclass[12pt]{article}
\usepackage[utf8]{inputenc}

\usepackage[title]{appendix}%

 \usepackage{natbib}
 \usepackage{nccmath}
  \usepackage[T1]{fontenc}
 \usepackage[english]{babel}
 \usepackage{mathtools, bm}
 \usepackage{amssymb, bm}
 \usepackage{stmaryrd} 
 \usepackage{enumitem}
 \usepackage{array}
 \usepackage{pgfgantt}
\usepackage{algorithm}
\usepackage{algpseudocode}
 \usepackage{layout}
 \usepackage{pslatex}
 \usepackage{pstricks-add}
 \usepackage{lmodern}
 \usepackage{listings}
 \usepackage{color}
 \usepackage[pdfborder={0 0 0}]{hyperref}
 \usepackage[babel=true]{csquotes}
 \usepackage{thmtools, varioref}
 \usepackage{lmodern}
 \usepackage{tikz}
 \usepackage{setspace}
 \usepackage{amsmath}
 \usepackage{dsfont}
 \usepackage{slashed} 
\usepackage{pdfpages}
 \usepackage{mathrsfs} 
 \usepackage{geometry}
\usepackage[standard]{ntheorem}
\usepackage{tcolorbox}
\usepackage{amsmath,amsfonts}
\usepackage{caption}
\usepackage{subcaption}
\usepackage{times,color,dsfont}
\usepackage{bm}
\usepackage[algo2e]{algorithm2e} \usepackage{diagbox}
\usepackage{multirow}
\usepackage{graphicx}
\newtheorem{assumption}{Assumption}

\newcommand{\E}{\mathrm{E}}

\newcommand{\emv}[1]{\hat{#1}_N}
\newcommand{\emvr}[1]{\tilde{#1}_N}
\newcommand{\emvB}[1]{{#1_N^*}}

\newcommand{\n}{N}
\newcommand{\J}{J_i}

\newcommand{\temps}{t}
\newcommand{\df}{p}
\newcommand{\dt}{r}
\newcommand{\bbr}{\mathbb{R}}
\newcommand{\Id}[1]{I_{#1}}
\newcommand{\obs}{y}

\newcommand{\meanrandom}{\beta}

\newcommand{\pr}{\textnormal{pr}}
\newcommand{\Y}{\obs_{1:\n}}

\newcommand{\lnb}{l(\theta;\obs_{1:\n}^*)}
\newcommand{\lnz}{l(\theta ; \obs_{1:\n})}
\newcommand{\va}{\textnormal{var}}
\newcommand{\lrt}{\textsc{lrt}(\Y)}
\newcommand{\lrtinf}{\textsc{lrt}_\infty}
\newcommand{\lrtb}{\textsc{lrt}(\obs^{*}_{1:\n})}
\newcommand{\lrtbb}{\textsc{lrt}(\obs^{*,b}_{1:\n})}
\newcommand{\lz}{l(\theta ; \obs_1)}
\newcommand{\cov}{x}
\newcommand{\diag}{\textnormal{diag}}

\newcommand{\fv}[2]{f_i(#1;#2)}
\newcommand{\fc}[1]{f_i(y_i;\xi_i,#1)}

\newcommand{\f}{f_i(\obs_i;\theta)}
\newcommand{\F}{f(\obs;\theta)}
\newcommand{\Fc}[1]{f(\obs;#1,\theta)}
\newcommand{\Fp}{f(\obs;\theta')}
\newcommand{\Fcp}[1]{f(\obs;#1,\theta')}

\newcommand{\diff}[1]{(#1-#1_0)}

\newcommand{\diiii}[4]{\sum_{#1,#2,#3,#4=1,...,d_\lambda}\lambda_{#1}\lambda_{#2}\lambda_{#3}\lambda_{#4}\frac{\partial^4 l(\theta_0;\obs_{1:\n})}{\partial\lambda_{#1}\partial\lambda_{#2}\partial\lambda_{#3}\partial\lambda_{#4}}}

\newcommand{\diiiid}[4]{\sum_{#1,#2,#3,#4=1,...,d_\delta}\delta_{#1}\delta_{#2}\delta_{#3}\delta_{#4}\frac{\partial^4 l(\theta_0;\obs_{1:\n})}{\partial\delta_{#1}\partial\delta_{#2}\partial\delta_{#3}\partial\delta_{#4}}}

\newcommand{\diid}[2]{\sum_{#1,#2=1,...,d_\delta}\delta_{#1} \delta_{#2}\frac{\partial^2 l(\theta_0;\obs_{1:\n})}{\partial\delta_{#1}\partial\delta_{#2}}  }
\newcommand{\dii}[2]{\sum_{#1,#2=1,...,d_\lambda}\lambda_{#1} \lambda_{#2}\frac{\partial^2 l(\theta_0;\obs_{1:\n})}{\partial\lambda_{#1}\partial\lambda_{#2}}  }
\newcommand{\dpsi}{\nabla_{\psi}l(\theta_0;\obs_{1:\n})}
\newcommand{\dpsii}{\nabla_{\psi}^2 l(\theta_0;\obs_{1:\n})}

\title{Bootstrap test procedure for variance components in nonlinear mixed effects models in the presence of nuisance parameters and a singular Fisher Information Matrix}
\author{T. GUEDON, C. BAEY, E. KUHN}
\date{}

\begin{document}
\setstretch{1.5}

\maketitle

\begin{abstract}
    We examine the problem of variance components testing in general mixed effects models using the likelihood ratio test. We account for the presence of nuisance parameters, i.e. the fact that some untested variances might also be equal to zero. Two main issues arise in this context leading to a non regular setting. First, under the null hypothesis the true parameter value lies on the boundary of the parameter space. Moreover, due to the presence of nuisance parameters the exact location of these boundary points is not known, which prevents the use of classical asymptotic theory of maximum likelihood estimation. Then, in the specific context of nonlinear mixed-effects models, the Fisher information matrix is singular at the true parameter value. 
    We address these two points by proposing a shrinked parametric bootstrap procedure, which is straightforward to apply even for nonlinear models. We show that the procedure is consistent, solving both the boundary and the singularity issues, and we provide a verifiable criterion for the applicability of our theoretical results. 
    We show through a simulation study that, compared to the asymptotic approach, our procedure has a better small sample performance and is more robust to the presence of nuisance parameters. A real data application is also provided. 
    
\end{abstract}

\noindent\textit{keywords : }
    Mixed effects model; Variance components testing; Parameters on the boundary; Bootstrap ; Nuisance Parameter; Singularity.

\tableofcontents

\section{Introduction}

Mixed effects models are a powerful statistical tool to model longitudinal studies with repeated measurements or data with an underlying unknown latent structure as hierarchical data. There are many fields of applications, e.g. pharmacokinetic-pharmacodynamic \citep{bonate2011}, medicine \citep{brown2015}, agriculture \citep{zhou2022}, ecology \citep{bolker2009}, psychology \citep{meteyard2020} or educational and social sciences \citep{gordon2019}. These models allow to take into account two types of variabilities, between different  individuals in a population and between several measurements made on the same individual, also called inter and intra variabilities. These are modeled by two types of effects: on the one hand, random effects that vary from one individual to another, and on the other hand, fixed effects, common to all individuals in the population \citep{pinheiro2006mixed,davidian2017nonlinear}. 

From a modeling point of view, being able to distinguish among all effects those that can be modeled as fixed effects would allow one to reduce the number of model parameters. This would also help to better identify the processes that are the cause of the variability observed in the population. Two main approaches have been developed to tackle this task. On the one hand, some authors suggested methods based on variable selection, using a Bayesian procedure as in \cite{chen2003random} or a penalized likelihood approach as in \cite{ibrahim2011fixed} or \cite{groll2014variable}. Specific selection criteria for mixed-effects models were also developed by \cite{vaida2005conditional,gurka2006selecting} and \cite{delattre2014note}. On the other hand, other authors focused on hypothesis testing for the nullity of some variance components of the random effects. 

Such a test is equivalent to comparing two nested models, and standard tools to address this question include the likelihood ratio, the score and the Wald tests statistics \citep{van2000asymptotic}.
However two issues arise when testing the nullity of variance components, that prevent from using the usual asymptotic results of \cite{wilks}. The first issue results from the true value of the variance parameter lying on the boundary of the parameter space, while the second is due to the singularity of the Fisher information matrix. 

In the specific context of mixed effects models, \cite{Crai04} derived the exact distribution of the likelihood ratio test statistic in a linear model with one random effect. \cite{stramlee94} derived an asymptotic likelihood ratio test for linear models in some specific cases. \cite{lin1997variance} proposed a score test in several specific cases in generalized linear mixed effects models.  \cite{wood2013simple} proposed a way to treat variance components testing in generalized models using the linear case machinery.  \cite{baey2019asymptotic} derived the asymptotic distribution of the likelihood ratio test statistic for testing that any subset of the variances of the random effects is null.  However, for applicability, these references assume that the untested parameters do not lie on the boundary of the parameter space.

In a more general framework, as far as the boundary issue is concerned, several authors studied the asymptotic of the likelihood ratio test statistic in this context. \cite{Cher54}, \cite{Chant74}, \cite{self1987asymptotic} and \cite{geyer1994asymptotics} derived the asymptotic distribution of the likelihood ratio test statistic in specific cases. \cite{andrews1999estimation}, \cite{silvapulle2005constrained} gave a more general way of dealing with hypothesis testing when the true parameter is not constrained to be an interior point of the ambient space.  When this is the case, the asymptotic distribution of the likelihood ratio test statistic is intractable as it depends on the unknown location of these nuisance parameters on the boundary \citep{self1987asymptotic}. Therefore, procedures that do not involve the asymptotic distribution of the test statistic can be preferred. In particular, resampling methods, such as those based on the bootstrap or permutations, are powerful tools to address this issue. In addition, these methods are usually more robust in small samples context. Dealing with variance components testing,\cite{sinha2009bootstrap} proposed a bootstrap-based score test in the context of generalized linear mixed models with one single random effect. \cite{Drik13} proposed a permutation-based test for any subset of the covariance matrix of the random effects in linear mixed models. The latter method is very easy to use in practice but is restricted to the context of linear models. The former requires the computation of the Fisher information matrix, which can be heavy  in practice, especially in the context of nonlinear models. Moreover, the presence of nuisance parameters on the boundary of the parameter space is not considered in the aforementioned works, even though it can be a source of inconsistency for the bootstrap procedures.  Indeed, as discussed in \cite{beran1997diagnosing}, and highlighted in  \cite{andrews2000inconsistency}, the bootstrap is known to be inconsistent when the true parameter value is a boundary point. 
When estimating the expectation of a Gaussian distribution, restricted to be nonnegative, \cite{andrews2000inconsistency} proposed 
a parametric procedure that shrinks the parameter used to generate the bootstrap data near the boundary. Following this idea, \cite{cavaliere2020bootstrap} proposed a more general parametric bootstrap test procedure based on the Likelihood Ratio Test statistic with parameters lying on the boundary. Their method consists in shrinking the bootstrap parameter in order to accelerate its rate of convergence toward the boundary. 

The second issue is the singularity of the Fisher information matrix that arises specifically in the context of mixed effect models,  as discussed in \cite{ekvall2021confidence}.This phenomenon is studied in \cite{rotnitzky2000likelihood} when the rank of the Fisher Information Matrix is full minus one.
Following the development of \cite{testinghiroyuki2012} to derive the new asymptotic distribution of the likelihood ratio test statistic, we show that this singularity issue is another source of inconsistency of the bootstrap procedure.

In this work we propose a shrinked parametric bootstrap test procedure for variance components in nonlinear mixed effects models that addresses the two issues mentioned above. We show that given an appropriate choice of the bootstrap parameter, the procedure is consistent as the number of individuals grows to infinity.  Our contribution is twofold: first, our procedure can be applied to linear, generalized linear and nonlinear models, and second, it takes into account the presence of nuisance parameters at unknown locations. We also provide a verifiable criterion to check the required regularity conditions. Finally, we illustrate our results on simulated and real data, exhibiting the good finite sample properties of the procedure and its applicability in practice.


\section{Proposed methodology}

\subsection{Mixed effects models}\label{sec:model}
Consider $\n$ individuals each measured $\J<J$ times, where $\n$ and $\J$ are nonnegative integers. We denote by $\obs_{ij}$ ($i=1,...,\n; j=1,...,\J$) the $j$th observation of the $i$th individual and we define $\obs_i=(\obs_{i1},\dots, \obs_{i\J})$ and $\obs_{1:\n} = (y_1^T,\dots,y_\n^T)$. In the sequel, $\mathcal{L}_\df^+$ denotes the space of lower triangular matrices of size $\df \times \df$ with positive diagonal coefficients, $\mathbb{S}^p_+$ denotes the space of symmetric, positive semi-definite $p \times p$ matrices, $\Id{\df}$ is the identity matrix of size $\df\times \df$, $[A]_{ij}$ is the element on the $i$th line and $j$th column of matrix $A$, and $\mathcal{\n}(\mu,V)$ denotes the multivariate Gaussian distribution with expectation $\mu\in\bbr^\df$ and covariance matrix $V$ of size $\df\times\df$. We consider the following nonlinear mixed effects model
\begin{equation}\label{eq:model}
    \left\{ 
    \begin{array}{ll}
          \obs_{ij} &= g(\cov_{ij},\beta,\Lambda\xi_i)+\varepsilon_{ij}\quad \varepsilon_{ij} \sim \mathcal{\n}(0,\sigma^2) \\
         \xi_i & \sim\mathcal{\n}\left(0,\Id{\df} \right) \\
    \end{array} \right. ,
\end{equation}
where $(\xi_i)_{i=1,\dots,\n}$ and $(\varepsilon_{ij})_{i=1,..,\n,j=1,...,\J}$ are mutually independent random variables, $g$ is a known nonlinear function, $\cov_{ij}$ gathers all the covariates of the $j$th observation of the $i$th individual, $\beta \in \bbr^{b}$ is the vector of fixed effects, $\Lambda \in \mathcal{L}_\df^+$ is a scaling parameter for random effect $\xi_i$, and $\sigma^2$ is the positive noise variance. 
The main advantage of this formulation for a nonlinear mixed effects model is that the random effect distribution is parameter-free, which will be particularly well adapted for the theoretical analysis of the proposed procedure.

\begin{remark}\label{rem:chol}
 The covariance matrix of the scaled random effect $b_i=\Lambda\xi_i$ is equal to $\Gamma = \Lambda\Lambda^T$ which is positive semi-definite. Therefore, a natural choice for $\Lambda$ is the lower triangular matrix in the Cholesky decomposition of the scaled random effects covariance matrix.
This reparametrization is used for instance in \cite{chen2003random}. When $\Gamma$ is positive definite, the Cholesky decomposition and hence the matrix $\Lambda$, is uniquely defined.
When $\Gamma$ is positive semi-definite, there exists a permutation of rows and columns of $\Gamma$ such that the permuted matrix has a unique Cholesky decomposition \citep{higham1990analysis}. This permutation orders the rows and columns of $\Gamma$ so that $\Lambda$ is lower triangular, with some diagonal blocks that can be equal to 0. 
\end{remark}

\begin{remark}
     The definition of model \eqref{eq:model} is slightly more general than the usual terminology of mixed effects models \citep[p. 306]{pinheiro2006mixed} that defines $\obs_{ij}  = g(v_{ij},\phi_i) + \varepsilon_{ij}$, with  $\phi_i = A_{ij}\beta+B_{ij}b_i$ the $i$th individual parameter, $\beta$ the vector of fixed effects associated with random effect $b_i\sim\mathcal{\n}(0,\Gamma)$, and where $v_{ij}$, $A_{ij}$ and $B_{ij}$ are known covariates. Model \eqref{eq:model} covers this definition by taking $x_{ij}=(v_{ij},B_{ij},A_{ij})$ and $b_i = \Lambda \xi_i$. For example, when considering a linear mixed effects model one writes $$ g(x_{ij},\beta,\Lambda\xi_i)  = x_{ij}^T(\beta+\Lambda\xi_i)$$ 
    where $x_{ij}$ are known covariates, $\beta $ is a unknown vector of fixed effects parameters, $\Lambda$ is an unknown scaling parameter and $\xi_i$ is the random effect.  
      One can also consider the logistic growth model \citep{pinheiro2006mixed} given by: $$g(x_{ij},\beta,\Lambda\xi_i)= \frac{\beta_1+ \lambda_1 b_{i1}}{1+\exp\left\{-\frac{x_{ij}-(\beta_2+ \lambda_2 b_{i2})}{\beta_3+ \lambda_3 b_{i3}}\right\}}$$
    where $\Lambda=\diag(\lambda_1,\lambda_2,\lambda_3)$ is supposed diagonal and $(x_{ij})_{j=1,...,J_i}$ are the times of measurements of individual $i$.
         A more detailed development of the general differences between those two parameterizations is given in section \eqref{sec:diffparam} of the supplementary material.
\end{remark}

Let us denote by $\theta = (\meanrandom, \Lambda, \sigma^2)$ the unknown vector of model parameters taking values in $\Theta$, by $\fv{\cdot}{\theta}$ the density of the $i$th individual response $\obs_i$ given a parameter $\theta\in\Theta$, by $\fc{\theta}$ the conditional density of $\obs_i$ given the random effect $\xi_i$ and a parameter $\theta$, and by $\pi_\df(\cdot)$ the density of the  $\df$-dimensional standard Gaussian density.
With these notations we can define the log-likelihood of the model given the $\n$-sample $\obs_{1:\n}$ by
\begin{equation}\label{eq:likelihood}
    l(\theta ; \obs_{1:\n}) =\log\{L_\theta(\obs_{1:\n})\}= \log\{\prod_{i=1}^{\n}\f\}= \sum_{i=1}^{\n}\log\{\int \fc\theta \pi_{p}(\xi_i)d\xi_i\}\end{equation}
We consider the marginal likelihood defined as the complete likelihood integrated over the distribution of the random effects, since the random effects $\xi_i$ are unobserved.
We recall that contrary to the usual formulation of nonlinear mixed effects models, where the random effects are defined as the scaled version $b_i$, the one considered in \eqref{eq:model} has the advantage to lead to a parameter-free distribution for the random effects $\xi_i$. Indeed, with the former definition, the distribution of the latent variables depends on $\Lambda$, and is not defined on the entire parameter space since we only constrain $\Gamma = \Lambda\Lambda^T$ to be positive semi-definite. When dealing with linear models, since the variance of the random effects adds up with the noise variance, the fact that some diagonal components in $\Gamma$ are null is not an issue. However, the change of variables $b_i = \Lambda\xi_i$ is a $\mathcal{C}^1$ diffeomorphism if and only if the diagonal coefficients of $\Lambda$ are strictly nonnegative.  Without this assumption the two parametrizations are no longer equivalent as illustrated in the supplementary material (see section \ref{sec:diffparam}). Our parametrization is similar to the so-called reparametrization trick proposed by \cite{kingma2014}  to train variational autoencoders with back-propagation. 

\subsection{Variance components testing}\label{sec:hyp}
 
Let $\dt \in \{1,\dots, \df\}$ be the number of variances to be tested. Without loss of generality we assume that we test the nullity of the last $\dt$ variances in $\Gamma = \Lambda\Lambda^T$. Therefore let us consider the following block matrix notation 

 \begin{equation*}
\Lambda= \left(\begin{array}{c|c}
\Lambda_1 & 0_{(\df-\dt)\times\dt} \\
\hline
\Lambda_{12} & \Lambda_2 
\end{array}\right),
\end{equation*}
where $\Lambda_1 \in \mathcal{L}_{\df-\dt}^+$, $\Lambda_2 \in \mathcal{L}_{\dt}^+$ and $\Lambda_{12} \in \mathcal{M}_{\dt\times (\df-\dt)}\left(\mathbb{R}\right)$. 
We write $\theta_0$ the true parameter on which we consider the following test : 

\begin{equation}\label{eq:test}
H_0 :  \theta_0\in\Theta_0\hspace{0.3cm}\text{against}\hspace{0.3cm}H_1 :  \theta_0\in\Theta,
\end{equation}
where
\begin{align*}
    \Theta_0 & = \{ \theta \in \bbr^q \mid \beta \in \bbr^b, \Lambda_1 \in \mathcal{L}_{p-r}^{+}, \Lambda_2=0, \Lambda_{12} =0, \sigma^2\in  \bbr_*^+ \},\\
    \Theta & = \{ \theta \in \bbr^q \mid \beta \in \bbr^b, \Lambda \in \mathcal{L}_{p}^{+},  \sigma^2\in  \bbr_*^+ \}.
\end{align*}

\begin{remark}\label{rem:nuisance}
    We do not impose the diagonal of $\Lambda_1$  to be strictly non-negative, which enables the case where some untested variances of the scaled random effects are in fact equal to zero. This will be discussed in more details in section \eqref{sec:theo_intro} with the definition of nuisance parameters.
\end{remark}

The likelihood ratio test statistic is defined as 

\begin{equation*}
\lrt = 2\left\{\underset{\theta\in\Theta}{\sup}\quad l(\theta;\obs_{1:\n}) - \underset{\theta\in\Theta_0}{\sup}\quad l(\theta;\obs_{1:\n})\right\}.
\end{equation*}
In order to test \eqref{eq:test} with a nominal level $0<\alpha<1$, we define the rejection region as $R_\alpha =\{\lrt\geq q_\alpha\}$ with $q_\alpha$ being the $(1-{\alpha})$th quantile of the distribution of $\lrt$. Unfortunately, this distribution is often intractable. 

In the following section, we detail the proposed shrinked parametric bootstrap procedure to test \eqref{eq:test}.

\subsection{Testing procedure}\label{sec:bootproc}

Following the lines of \cite{cavaliere2020bootstrap}, we propose a parametric bootstrap procedure using a bootstrap parameter $\emvB{\theta}$ and $B\in\mathds{N}^*$ bootstrap replications to test \eqref{eq:test} with a type I error $0<\alpha<1$. As introduced in remark \ref{rem:nuisance} and then detailed in section \eqref{sec:theo_intro}, some untested variances, at unknown locations, can be null. Therefore using $\emv{\theta}=(\emv{\beta},\emv{\Lambda},\emv{\sigma}^2)$ the maximum likelihood estimator as a bootstrap parameter over $\Theta$ would fail to asymptotically mimic the true distribution of the likelihood ratio test statistic. Indeed there are elements in $\Lambda_0$ which are supposed to be zero, but that are non null in $\emv{\Lambda}$.Since we require that $\emvB{\theta}\in\Theta_0$ we can choose $\emvB{\theta}$ to be the unrestricted maximum likelihood estimator projected on $\Theta_0$ or the restricted one. Furthermore, we use a shrinking parameter $c_\n$ to fix to zero the untested components of $\Lambda$ that are smaller than $c_\n$. The proposed algorithm is described in algorithm \ref{alg:boot}, and the theoretical justification of this shrinking procedure is described in section \ref{sec:theo}.

\begin{algorithm}\caption{Shrinked parametric bootstrap for variance components testing}\label{alg:boot}
\begin{tabbing}
    \qquad \enspace Input:  $c_\n>0$, $B\in\mathds{\n}^*$, $0<\alpha<1$ \\
    \qquad \enspace Set $\emvB{\beta} = \emv{\beta}$, $\emvB{\Lambda} = \emv{\Lambda}$, and $\emvB{\sigma}^2= \emv{\sigma}^2$\\ 
    \qquad \enspace Set  $\Lambda_{2,\n}^*=\Lambda_{12,\n}^*=0$ \\ 
    \qquad \enspace Set $[\Lambda_{1,\n}^*]_{mn}=[\hat{\Lambda}_{1,\n}]_{mn}\mathds{1}_{[\hat{\Lambda}_{1,\n}]_{mn}>c_\n}$ \\
    \qquad \enspace For $b=1,...,B $ \\
        \qquad \qquad For $i=1,...,\n$, draw independently $\varepsilon_i^{*,b} \sim \mathcal{\n}(0,\emvB{\sigma}^2\Id{\J})$ and $\xi_i^{*,b} \sim \mathcal{\n}(0,\Id{\df})$ \\ 
        \qquad \qquad Build the $i$th value of the $b$th bootstrap sample $\obs_i^{*,b}=g(x_i,\emvB{\beta},\emvB{\Lambda}\xi_i^{*,b}) + \varepsilon_i^{*,b}$ 
        \\
        \qquad \qquad  Compute the likelihood ratio statistic $\lrtbb$\\
    \qquad \enspace Compute the bootstrap $p$-value as $p_{boot} = \frac{1}{B} \sum_{b=1}^B \mathds{1}_{\lrtbb > \lrt}$ \\
    \qquad \enspace Reject $H_0$ if $p_{boot}<\alpha$  
\end{tabbing}
\end{algorithm}

The next section is dedicated to the asymptotic validity of this testing procedure. 






\section{Theoretical results}\label{sec:theo}

\subsection{Notations and theoretical setting}\label{sec:theo_intro}

In this section we are interested in the theoretical consistency of the bootstrap procedure presented in section \ref{sec:bootproc}. We consider the asymptotic as the number of individuals $\n$ grows to infinity, while the number of measurements per individual remains fixed and bounded by some value $J$. We denote by $\theta_0\in\Theta_0$ the true value of the parameter, such that the density of the response $\obs_{1:\n}$ is $L_{\theta_0}(\obs_{1:\n})$. We denote by $E\{T(\obs_{1:\n})\}$ the expectation of any measurable function $T$ of $\obs_{1:\n}$ if there is no confusion about the distribution of $\obs_{1:\n}$. Otherwise we specify $E_\theta\{T(\obs_{1:\n})\}$ to emphasize that the expectation is with respect to the density $L_\theta(\obs_{1:\n})$, for any $\theta\in\Theta$. As commonly used in the bootstrap literature, we denote by $X^*$ the bootstrap version of a random variable $X$. We write $E^*\{T(y_{1:\n}^*)\} = E_{\emvB{\theta}}\{T(y_{1:\n}^*)\mid \obs_{1:\n}\}$. Similarly, for any measurable subset $A$ we write $ \pr^*\{T(\obs_{1:\n}^*)\in A\} = E_{\emvB{\theta}}\{\mathds{1}_{T(\obs_{1:\n}^*)\in A}\mid \obs_{1:\n}\} $. 

We want to show that the proposed bootstrap procedure is asymptotically valid which means that $\lrtb$ converges weakly in probability to the same limiting distribution as the one of $\lrt$. More precisely, we want to show that, under some conditions, if there exists a random variable $\lrtinf$ such that $\lrt$ converges weakly to $\lrtinf$ then for every $t\in\bbr$, as $\n\rightarrow +\infty$, it holds in probability that
\begin{equation}\label{eq:lrtboot}
    \pr^*\{\lrtb\leq t \} \longrightarrow \pr(\lrtinf\leq t).
\end{equation}

 We also use the notations $o_p(1)$ and $O_p(1)$ for random sequences that respectively converge toward zero and are bounded in probability. More generally, this notation is used to compare two random sequences, using the definition of \citet[section 2.2]{van2000asymptotic}. We also use their bootstrap versions $o_{p^*}$ and $O_{p^*}$ defined as follows: for a random quantity $X_\n^*$ computed on the bootstrap data, $X_\n^*=o_{p^*}(1)$ means that for any $\varepsilon>0$, $\pr^*(X_\n^*>\varepsilon)\rightarrow 0$ in probability as $\n\rightarrow +\infty$. Similarly $X_\n^*=O_{p^*}(1)$ means that for any $\varepsilon>0$ there exists a real $M>0$ and an integer $\n_0$ such that for all $\n>\n_0$, the event $\{\pr^*(\|X_\n^*\|>M)<\varepsilon\}$ is arbitrary close to one in probability.

We now formalize what we call nuisance parameters. We suppose that, in addition to the last $r$ tested variances, $m$ untested variances are null. Without loss of generality we suppose that the last $m+r$ variances of the individual parameters are null therefore $\Lambda_0$ is of the form
\[
    \Lambda_0 = 
             \left(\begin{array}{c|c|c} 
                    \Lambda_1^{nonuis} & 0_{(\df-\dt-m)\times m}& 0_{(\df-\dt-m)\times\dt} \\
                    \hline
                    \Lambda_{12}^{nuis} & \Lambda_1^{nuis} & 0_{m\times \dt}\\
                    \hline
                    \Lambda_{12,1} & \Lambda_{12,2} &  \Lambda_2\\
                \end{array}\right)
            = \left(\begin{array}{c|c|c}
                    \Lambda_1^{nonuis}  & 0_{(\df-\dt-m)\times m}& 0_{(\df-\dt-m)\times\dt} \\
                    \hline
                    0_{m\times (\df-\dt-m)} & 0_{m\times m} & 0_{m\times \dt}\\
                    \hline
                    0_{\dt\times (\df-\dt-m)}  & 0_{\dt\times m} &  0_{\dt\times \dt}\\
                \end{array}\right).
\]
It is important to notice that in real life applications the $m$ rows inducing nuisance parameters are located at unknown positions in matrix $\Lambda$, and that the remaining $\df-m-r$ variances are strictly non-negative which is equivalent to the diagonal coefficients of $\Lambda_1^{nonuis}$ being strictly non-negative.

Following \cite{self1987asymptotic} and \cite{cavaliere2020bootstrap}, we now split the parameter as $\theta = (\psi, \delta, \lambda)$, where $\lambda$ stands for all the coefficients of $\Lambda_2$ and $\Lambda_{12,2}$, $\delta$ represents the coefficients in $\Lambda_1^{nuis}$ and $\psi$ gathers all the remaining parameters. The dimension of $\lambda$ is $d_\lambda=r(r+1)/2 + r(p-r-m)$, the dimension of $\delta$ is $d_\delta=m(m+1)/2+r\times m$ and the dimension of $\psi$ is $d_\psi=d_\theta-d_\lambda-d_\delta$. Moreover, $\theta_0=(\psi_0,\delta_0,\lambda_0)=(\psi_0,0_{d_\delta}, 0_{d_\lambda})$. Before introducing our results we first state a set of general conditions on the model that will be required in this work. 

\begin{assumption}\label{ass:regularity1}
$(i) \Theta$ is compact, $(ii)$  the model is identifiable, $(iii)$ for all $i\in\mathds{\n}, \obs\in\mathds{R}^\J$, $\xi\in\mathds{R}^\df$ the conditional likelihood $\theta\mapsto \fv{\obs}{\xi,\theta}$ is 4-times differentiable on the interior of $\Theta$, and directional derivatives exist on the boundary, $iv)$ each partial derivative of $\theta\mapsto \fv{\obs}{\xi,\theta}$ is bounded by a positive function which does not depend on $\theta$ and is integrable with respect to the distribution of the random effects.
\end{assumption}

\begin{remark}
     The compactness assumption  is not verified for $\Theta$. However in practice it only requires that $\sigma^2\geq\rho$ for some non-negative number $\rho$ and that each component of $\theta$ is upper and lower bounded, which is reasonable in real data applications.Assumption $(ii)$ is usual in the context of estimation theory. Assumption $(iii)$ is needed to perform a Taylor expansion of the log likelihood and $(iv)$ is needed to differentiate under the integral sign in \eqref{eq:likelihood}. These assumptions are  discussed in section \eqref{sec:verifG}.
\end{remark}

The following proposition induces that if the Fisher Information Matrix exists, it will present blocks equal to zero, and will therefore be singular. This result extends the one of \cite{rotnitzky2000likelihood} stated when the rank of the Fisher Information Matrix is full minus $1$ to more general settings where the rank of the Fisher Information Matrix is full minus $d_{\lambda}+d_\delta$.

\begin{proposition}\label{lem:singularity}
Under assumption \eqref{ass:regularity1}, for $k=0, 1$, for all $i\in\mathds{\n}$ and for all $y\in \bbr^{\J}$, $\nabla^{2k+1}_\delta \log\{\fv{\obs}{\theta_0}\} = 0_{d_\delta^{2k+1}}$ and $\nabla^{2k+1}_\lambda \log\{\fv{\obs}{\theta_0}\}=0_{d_\lambda^{2k+1}}$. In particular, $\va  \{\nabla_\delta\lnz\} = 0_{d_\delta\times d_\delta}$ and $\va \{\nabla_\lambda \lnz\} = 0_{d_\lambda\times d_\lambda}$.
\end{proposition}

\begin{remark}
    If $\theta\mapsto\lnz$ admits higher order derivatives, the first part of Proposition \ref{lem:singularity} is true for every odds order derivatives. This comes from the null odds moments of the standard normal distribution of the random effects.
\end{remark}

\begin{remark}
     As shown in the proof of proposition \ref{lem:singularity}, in section \ref{app:prop1proof} of the Appendix,  by considering the $k$th column $[\Lambda]_{.k}=([\Lambda]_{1k},...,[\Lambda]_{\df k})^T$ of $\Lambda$, for all $j=1,...,\df$, $\partial \lnz/\partial [\Lambda]_{jk}|_{[\Lambda]_{.k}=0_\df}=0$. That explains why the coefficients of $\Lambda_{12,2}$ are part of the definition of $\lambda$.
\end{remark}

\subsection{Consistency of the bootstrap procedure in the identically distributed setting}

We first deal with the simpler identically distributed case. In model \eqref{eq:model} it corresponds to the case where $(\cov_{ij})_{j=1,...,\J}$ are common to every individual $i$. The next section is devoted to extending the results to the non identically distributed setting presented before.

Before studying the consistency of the test procedure, we first need to ensure the consistency of the restricted (respectively unrestricted) maximum likelihood estimator, i.e. computed over $\Theta_0$ (respectively $\Theta$). We first state the regularity conditions required for the asymptotic theory that follows.
\begin{assumption}\label{ass:regularity 0-4} For every $k,l,s,t=1,\ldots,d_\theta$ and for every $i=1, \ldots , N$:
\begin{fleqn}
\begin{alignat*}{1}
(i)\quad\underset{\theta'\in\Theta}{\sup} \E_{\theta'}\{\underset{\theta\in\Theta}{\sup}&|\log f(\obs_i;\theta)|^2\}<+\infty \\
(ii)\quad\underset{\theta'\in\Theta}{\sup} \E_{\theta'}\{\underset{\theta\in\Theta}{\sup}&\left|\frac{\partial \log f(\obs_i;\theta)}{\partial \theta_k}\right|^3\}<+\infty \\
(iii)\quad\underset{\theta'\in\Theta}{\sup} \E_{\theta'}\{\underset{\theta\in\Theta}{\sup}&\left|\frac{\partial^2 \log f(\obs_i;\theta)}{\partial \theta_k\partial\theta_l}\right|^3\}<+\infty \\
(iv)\quad \underset{\theta'\in\Theta}{\sup} \E_{\theta'}\{\underset{\theta\in\Theta}{\sup}&\left|\frac{\partial^3 \log f(\obs_i;\theta)}{\partial \theta_k\partial\theta_l\partial\theta_s}\right|^2\}<+\infty \\
(v)\quad\underset{\theta'\in\Theta}{\sup} \E_{\theta'}\{\underset{\theta\in\Theta}{\sup}&\left|\frac{\partial^4 \log f(\obs_i;\theta)}{\partial \theta_k\partial\theta_l\partial\theta_s\partial\theta_t}\right|^2\}<+\infty 
\end{alignat*}
\end{fleqn}

\end{assumption}

Assumption \eqref{ass:regularity 0-4} $(i)$ is needed to ensure the consistency of the maximum likelihood estimators. Indeed it enables to derive a uniform law of large numbers. Assumptions $(ii)$ and $(iii)$ are similar to assumption (N8') in \cite{hoadley1971asymptotic}. It is required to apply a central limit theorem to the score function, and the pseudo score function $\tilde{S}_\n(\theta)$ that appears in the quadratic expansion (see equation \eqref{eq:quadratic2} in the Appendix). Assumptions $(iv)$ and $(v)$ are needed to control the rest of the quadratic approximation. All the suprema are needed to control the consistency of the bootstrap distributions. 

We now derive the consistency of the maximum likelihood estimators, following the result of \cite{moran1971uniform}.

\begin{proposition}\label{prop:consist mle}
    Under assumptions \eqref{ass:regularity1}--\eqref{ass:regularity 0-4} $i)$ : 
    \begin{align*}
         \arg \underset{\theta\in\Theta}{\max}\quad \lnz & = \theta_0 +o_p(1)\\
    \arg \underset{\theta\in\Theta_0}{\max}\quad \lnz & = \theta_0 +o_p(1)
    \end{align*}
\end{proposition}

A natural choice for the bootstrap parameter $\emvB{\theta}$ is the maximum likelihood estimator. However the bootstrap fails in presence of the nuisance parameters summarized in vector $\delta$. This is why care must be taken when choosing $\emvB{\delta}$. To explain and solve this issue we first need to derive the speed of convergence of the maximum likelihood estimator. 

\begin{proposition}\label{prop:speed} 
Let $\emv{\theta}=(\emv{\psi},\emv{\delta},\emv{\lambda})$ and $\emvr{\theta}=(\emvr{\psi},\emvr{\delta},0_{d_\lambda})$ be respectively the unrestricted and restricted maximum likelihood estimators of $\theta$. 
   Under assumptions \eqref{ass:regularity1} and  \eqref{ass:regularity 0-4} 
        $(\emv{\psi},\emvr{\psi}) = O_p(\n^{-1/2}),
        (\emv{\delta},\emvr{\delta},\emv{\lambda}) = O_p(\n^{-1/4})$.
\end{proposition}

We emphasize that this result achieves the same rate of convergence as in \cite{rotnitzky2000likelihood}. However their result is not applicable here as the number of vanishing score components is greater than one, therefore the exact asymptotic distribution of the maximum likelihood estimators is unknown. The usual way to derive the asymptotic distribution of the likelihood ratio statistic is to consider a quadratic approximation of the log-likelihood around the true value of the parameter, based on a second-order Taylor expansion. However in our case, due to the vanishing score property stated in proposition \ref{lem:singularity}, this quadratic expansion is degenerate with respect to parameters $\delta$ and $\lambda$. Using a reparametrization of parameter $\theta$ as in \cite{testinghiroyuki2012}, we obtain a new quadratic approximation based on a higher-order expansion of the log-likelihood. We then apply results from \cite{andrews1999estimation} and \cite{silvapulle2005constrained} to obtain an explicit formula for $\lrtinf$. This new quadratic approximation involves a new matrix $\tilde{I}(\theta_0)$ which plays the role of the Fisher information matrix, and which is defined explicitly in equation \eqref{eq:quadratic2} of the supplementary material. This new matrix is no longer systematically degenerate, we can therefore state the usual assumption which must be verified case by case in real life applications.

\begin{assumption}\label{ass:Itilde}
$\tilde{I}(\theta_0)\succ 0$
\end{assumption}

     Matrix $\tilde{I}(\theta_0)$ is the asymptotic variance of a modified score function (see Remark \ref{rem:nonsing IN} in the Appendix), which ensures that it is positive semi-definite. This matrix depends on derivatives up to the fourth order of the log-likelihood, and is no longer degenerate. 

Before showing that the proposed test procedure is consistent, we first need to show that in the boostrap world, if the bootstrap parameter is consistent, the bootstrap maximum likelihood estimators are, conditionally to the data, consistent.

\begin{proposition}\label{prop:consistboot}
     Under assumptions \eqref{ass:regularity1}--\eqref{ass:regularity 0-4} $i)$, if $\emvB\theta$ the parameter used to generate the data is consistent, then : 
    \begin{align*}
         \arg \underset{\theta\in\Theta}{\max}\quad \lnb & = \theta_0 +o_{p^*}(1)\\
    \arg \underset{\theta\in\Theta_0}{\max}\quad \lnb & = \theta_0 +o_{p^*}(1).
    \end{align*}
\end{proposition}

We can now state the main result that guarantees the consistency of the bootstrap procedure.

\begin{theorem}\label{prop:LRTbootsasym}
Under assumptions (1)--(3), if $\emvB{\theta}$ is chosen such that $\emvB{\theta}\in\Theta_0$, $\emvB{\theta} = \theta_0+o_{p}(1)$ and $\n^{1/4}\emvB{\delta} = o_p(1)$ then as $\n\rightarrow +\infty$, it holds in probability that 
\begin{equation}
    \pr^*\{\lrtb\leq t \} \longrightarrow \pr(\lrtinf \leq t).
\end{equation}
where the expression of $\lrtinf $ is given in the Appendix.
\end{theorem}

A way of choosing $\emvB{\theta}$ that fulfills the hypotheses of theorem \ref{prop:LRTbootsasym} is to follow the idea of \cite{cavaliere2020bootstrap} and shrinks the parameter toward 0. However here the rate of convergence of the shrinking parameter $(c_\n)$ is not the same due to the singularity issue. The following lemma gives a procedure to choose $\emvB{\theta}$ and justify the way it is chosen in algorithm \ref{alg:boot}.

\begin{proposition}\label{lem:shrink}
    Let $(c_\n)_{\n\in\mathds{\n}}$ be a sequence such that $\underset{\n\rightarrow +\infty}{\lim} c_\n=0$ and $\underset{\n\rightarrow +\infty}{\lim}\n^{\frac{1}{4}} c_\n = + \infty$. Let $\hat{\theta}_\n = (\hat{\psi}_\n, \hat{\delta}_\n, \hat{\lambda}_\n)$ be a maximum likelihood estimator (restricted or not) of $\theta_0=(\psi_0,0_{d_\delta}, 0_{\delta_\lambda})$. 
    Under assumptions \eqref{ass:regularity1}--\eqref{ass:Itilde}, by choosing $\emvB{\theta} = (\emvB{\psi}, \emvB{\delta},\emvB{\lambda})$ such that: 
     $\forall k=1,..,d_\psi$ $\psi^*_{\n,k} = \hat{\psi}_{\n,k}\ \mathds{1}(\hat{\psi}_{\n,k}>c_\n)$, $\forall k=1,..,d_\delta$ $\delta^*_{\n,k} = \hat{\delta}_{\n,k}\ \mathds{1}(\hat{\delta}_{\n,k}>c_\n)$ and 
 $\emvB{\lambda} = 0_{d_\lambda}$
    then, $\emvB{\theta}$ verifies the hypothesis of theorem \ref{prop:LRTbootsasym}.
\end{proposition}

As we do not know which parameters are part of $\delta$, it is important to deal with every potential nuisance parameters. This is why we also consider a shrinkage bootstrap parameter for $\psi$. In the proof of this proposition we show that the shrinkage does not change the limit of the estimate, but only speeds up its convergence toward $0$.

In addition to the boundary issue, the singularity is another source of inconsistency for the bootstrap procedure. As highlighted in the proof of theorem \eqref{prop:LRTbootsasym}, this inconsistency comes from the polluting random variables due to the asymptotic distribution of $\n^\frac{1}{4}\emvB\delta$, that does not appear in the asymptotic distribution of $LRT_\infty$. The shrinkage enables to enforce that $\n^\frac{1}{4}\emvB\delta = o_p(1)$ and no longer $O_p(1)$.

\subsection{Extension to the non identically distributed setting}

 As in the previous section we first derive the consistency of the maximum likelihood estimator. To do so, we need the regularity required in assumption \eqref{ass:regularity 0-4} to hold uniformly over the different distributions of the individuals. 
 
  \begin{assumption}\label{ass:notiid}
  We suppose that assumptions \eqref{ass:regularity 0-4} $(i)$--$(v)$ hold uniformly over the different individuals $i\in\mathds{N}$.  

 \end{assumption}

In addition to that, as discussed in \cite{hoadley1971asymptotic} an additional assumption is required to ensure the unicity of the maximum of the asymptotic objective function. 
 
  \begin{assumption}\label{ass:not iid unicity}
  For every $\theta\neq\theta_0$ : 
          \[\underset{\n\rightarrow+\infty}{\lim} \frac{1}{\n}\sum_{i=1}^{\n}\E\left[\log\left\{\frac{f_i(\obs_i;\theta)}{f_i(\obs_i;\theta_0)}\right\}\right]<0\].
 \end{assumption}

\begin{proposition}\label{prop:mle not iid}
    Under assumptions \eqref{ass:regularity1}, \eqref{ass:notiid} and \eqref{ass:not iid unicity}, propositions \ref{prop:consist mle} and \ref{prop:consistboot} still hold in the non identically distributed case.
\end{proposition}

Following the same lines as in the last section the result of theorem \ref{prop:LRTbootsasym} still holds.

\begin{theorem}\label{prop:LRTbootsasym not iid}
    Under assumptions \eqref{ass:regularity1}--\eqref{ass:not iid unicity},  if $\emvB{\theta}$ is chosen such that $\emvB{\theta}\in\Theta_0$, $\emvB{\theta} = \theta_0+o_{p}(1)$, i.e. $\emvB{\lambda}=0$ and $\n^{\frac{1}{4}}\emvB{\delta} = o_p(1)$ then as $\n\rightarrow +\infty$, it holds in probability that
\begin{equation}
    \pr^*\{\lrtb\leq t \}- \pr(LRT_\n \leq t) = o_p(1).
\end{equation}

\end{theorem}

\subsection{Sufficient verifiable conditions for regularity assumptions}\label{sec:verifG}

Assumptions \eqref{ass:regularity 0-4} state regularity conditions on the model. These assumptions  can be straightforward to verify in some models (see details of the calculation in a linear  mixed effects model in Appendix \ref{sec:linear}). Nevertheless, in most of the nonlinear cases these assumptions are very difficult to check, in particular due to the non explicit integrated form of the likelihood in \eqref{eq:likelihood}. 
\cite{nie2006strong} proposed some verifiable conditions under which the maximum likelihood estimators in nonlinear mixed models is strongly consistent. However in his work he considered that the true parameter is an interior point of the parameter space, and that the Fisher information matrix is nonsingular. Furthermore in our context the conditions required are even more difficult to verify as we deal not only with maximum likelihood estimator consistency but also with likelihood ratio and bootstrap statistic consistency. 

We propose an analytical sufficient criterion for nonlinear mixed models  that only depends on the regularity of the known function $g$ in model \eqref{eq:model} when $g$ is nonlinear in the random effects $\xi_i$. 
We first state a regularity condition on the derivatives of $g$.

\begin{assumption}\label{eq:reg_g}
    For every $\xi$, $g$ is 4 times differentiable on $\Theta$, and for $k_1=0,...,4$ and $k_2\in\mathbb{N}$:
    \begin{equation}
        \underset{i\in\mathds{\n},j=1,...,\J}{\sup}\E\{\quad\underset{\theta\in\Theta}{\sup}\|\nabla^{k_1}_\theta g(\cov_{ij},\beta,\Lambda\xi)\|^{k_2}\}<+\infty,  \quad \xi \sim \mathcal{\n}(0,\Id\df).
    \end{equation}
\end{assumption}

\begin{remark}
    This assumption seems very strong but in practice it only requires that the derivatives of $g$ are not exponential in $\|\xi\|^2$ which is verified by almost every commonly used models.
\end{remark}

We now state the regularity condition on the function $g$, which is the proposed criterion to be verified case by case in real life applications.

\begin{assumption}\label{eq:reg_g2}  
    For every $\varepsilon>0$,  there exists a compact set $K\subset\mathds{R}^{\df}$ such that
\begin{equation}
 \forall \xi\in\mathds{R}^\df\backslash K\quad \underset{i\in\mathds{\n},,j=1,...,\J}{\sup}\underset{\theta\in\Theta}{\sup}\quad\frac{\|g(\cov_{ij},\beta,\Lambda\xi)\|}{\|\xi\|}\leq\varepsilon.
 \end{equation}
\end{assumption}

\begin{proposition}\label{prop:verif G} Suppose that assumption \eqref{ass:regularity1} holds, and that the function $g$ verifies assumption \eqref{eq:reg_g}--\eqref{eq:reg_g2}, then assumption \eqref{ass:notiid} is verified. 
\end{proposition}

\begin{remark}
Models with a bounded function $g$ verify this property for any compact set $K$ (see for example a common pharmacokinetic model presented in \citep{davidian2017nonlinear} and detailed in Appendix \ref{pharmaco}). Regarding  models with an unbounded function $g$, for example as the logistic growth model \citep{pinheiro2006mixed} used in the experiments section \ref{sec:simu}, one can verify case by case that  this criterion is satisfied (details of calculations are given in Appendix \ref{logistic_validity}).
\end{remark}

\section{Experiments}

\subsection{Simulation study}\label{sec:simu}

We denote by $\theta_0 = (\beta_0, \Lambda_0, \sigma^2_0)^T$ the true parameter used to generate the data. We use the notation $\beta_k$ for the $k$th component of the fixed effects vector $\beta$, and we write diag$(x_1,...,x_p)$ for a diagonal $p\times p$ matrix, with a diagonal being equal to $(x_1,...,x_p)^T$. When it is not explicitly written we consider diagonal matrices $\Lambda =\diag(\lambda_1, ... \lambda_\df)^T$. The same way we write $\xi_i=(\xi_{i1},...,\xi_{i\df})^T$ for the vector of random effects. 
We consider a linear and a nonlinear mixed effects models, with a varying number of random effects to account for the presence of nuisance parameters. Results were obtained using the \texttt{lme4} and \texttt{saemix} packages in R. Codes are available upon request from the first author.

We first consider the linear case. We denote by $m_1$ the linear model with two independent random effects, i.e. with $g(x_{ij},\beta,\Lambda \xi_i) = \beta_1 + \lambda_1 \xi_{i1} +(\beta_2+\lambda_2\xi_{i2})x_{ij}$. We set $\beta_0=(0,7)^T$, $\lambda_{01}=1.3,\lambda_{02}=0$. In this model, we consider the test $H_0: \lambda_2 = 0 $ against $H_1 : \lambda_2 \geq 0$. 
We then denote by $m_2$ the linear model with three independent random effects, i.e. with $g(x_{ij},\beta,\Lambda \xi_i) = \beta_1 + \lambda_1 \xi_{i1} +(\beta_2+\lambda_2\xi_{i2})x_{ij} +(\beta_3+\lambda_3\xi_{i3})x_{ij}^2 $.  We set $\beta_0=(0,7,3)^T$, $\lambda_{01}=1.3,\lambda_{02}=0$, $\lambda_{03} = 0$. In this model, we consider the test $H_0: \lambda_3 = 0 $ against $H_1 : \lambda_3 \geq 0$, so that in this simulation $\lambda_2$ is a nuisance parameter. For the choice of the shrinkage bootstrap parameter, we set $c_\n = a \n^{-\nu}$ with $a = 0.5$ and $\nu = 0.2,$ similarly to  \cite{cavaliere2020bootstrap}. This choice is motivated by the theoretical convergence assumptions on $c_\n$ in Proposition \ref{lem:shrink}. This parameter shrinks to zero the variances of the individual parameters $\beta_2+\lambda_2\xi_{i2}$ with a relative standard deviation lower than $4\%$. 
In both settings, we set $x_{ij} = j$, $J=5$ and $\sigma_0^2 = 1.5$.
Finally, we denote by $m_3$ the linear model with $\df=8$ random effects and a varying number $s$ of nuisance parameters, i.e. with $g(x_{ij},\beta,\Lambda \xi_i) = \sum_{k=1}^p x_{ijk}\lambda_k\xi_{ik}$. Here we set $\n = 40$, $\J=9$, $\sigma^2=1$  , and every untested variance to $1$. Finally we draw independently the covariates from a normal distribution with mean 2 and standard deviation 0.5. We want here to illustrate the effect of an increasing number of nuisance parameters on the performance of the test. We use three different values for the shrinkage parameter $c_N\in\{0; 0.24; 0.9\}$. We chose those values to consider three cases: first  $c_\n=0$ is equivalent to the parametric bootstrap procedure without shrinkage, then $c_\n=0.5 \times 40^{-0.2}\approx 0.24$ shrinks most of the nuisance parameters toward 0, and finally $c_\n=0.9$ shrinks systematically the nuisance parameters (as if we  were using the true model), but can also shrink some non-zero variances of the model. We consider the test $H_0: \lambda_1 = 0 $ against $H_1 : \lambda_1 \geq 0$.

Next, we consider the nonlinear logistic model with three random effects denoted by $m_4$, where 
\begin{equation}\label{eq:logisticmodel}
    g(x_{ij},\beta,\Lambda \xi_i) = \frac{\beta_1+\lambda_1\xi_{i1}}{1 + \exp \left\{ -\frac{x_{ij}-(\beta_2+\lambda_2\xi_{i2})}{\beta_3+\lambda_3\xi_{i3}} \right\}}.
\end{equation}
We set $\beta_0=(200,500, 150)^T$, $\lambda_{01}=\lambda_{02}=10$, $\lambda_{03} = 0$ and $\sigma_0^2 = 5^2$. We set $(x_{i1},...,x_{iJ}) = (50,  287.5,  525,  762, 1000, 1100, 1200, 1300, 1400, 1500)$ for all $i$. In this model, we consider the test $H_0: \lambda_3 = 0 $ against $H_1 : \lambda_3 \geq 0$.

First, we study the finite sample size properties of our procedure using models $m_1$, $m_2$ and $m_4$.  We compute the empirical levels by generating $K$ datasets under the null hypothesis as described in the previous paragraph, and by computing the proportion of these datasets for which we reject the null hypothesis, for a nominal level $\alpha$ in $\{0.01,0.05,0.10\}$ and a sample size $\n$ in $\{10,20,30,40,100\}$ for $m_1$, $\n$ in $\{20,30,40\}$ for $m_2$ and $\n=40$ for $m_4$. Results are given in tables \ref{tab:m1lvl}, \ref{tab:m2lvl} and \ref{tab:nonlin}.
We compare the empirical level of the test associated with our bootstrap procedure with those obtained using the asymptotic distribution which is a $0.5-0.5$ mixture between a Dirac distribution at zero and a chi--squared distribution with one degree of freedom \citep{baey2019asymptotic}. 
We observe that the empirical levels obtained with our bootstrap procedure are closer to the nominal ones than those obtained with the asymptotic procedure, and that good results are already obtained for small values of $\n$ in the linear case. As expected, our procedure exhibits better small sample size properties than the asymptotic procedure, both in the linear and the nonlinear cases. It is noteworthy to mention that the existing non-asymptotic test procedures such as the one proposed by \cite{Drik13} can not be used in the latter case since they rely on explicit expressions for the parameter estimates, hence requiring the linearity assumption. We also observe that the presence of nuisance parameters also deteriorate the asymptotic results. It is not a surprise as it modifies the true asymptotic distribution.  However we observe that the standard parametric bootstrap procedure is robust to the presence of a single nuisance parameter, and so the choice of $c_\n$ does not have a significant effect on this example. This must be due to the low number of nuisance parameters and the few number of parameters of the model.

\begin{table}
\def~{\hphantom{0}}
\caption{Empirical levels (expressed as percentages) of the test that one variance is null in a linear model with two independent random effects, for $K=5000$ simulated datasets and $B=500$ bootstrap replicates. The last column gives the maximal standard deviation value obtained in each row}{
\begin{tabular}{cccccccccccc}
 \multirow{2}{*}{\shortstack{Level \\ $\alpha$}} & \multicolumn{2}{c}{$\n=10$} & \multicolumn{2}{c}{$\n=20$} & \multicolumn{2}{c}{$\n=30$} & \multicolumn{2}{c}{$\n=40$} & \multicolumn{2}{c}{$\n=100$} & \multirow{2}{*}{\shortstack{max \\ sd}}  \\ 
   & boot & asym & boot & asym & boot & asym & boot & asym & boot & asym & \\
 1\% & 1.14 & 0.68 & 0.98 & 0.68 & 1.20 & 0.94 & 0.74 &  0.70 &   0.86 & 0.72 & 0.15 \\ 
   5\% & 5.20 & 3.64 & 5.22 & 3.82 & 5.74 & 4.30 & 4.86 & 3.94 & 5.26 & 4.50 & 0.33 \\ 
   10\% & 10.72  & 7.16 & 10.80  & 7.98 & 10.30 & 8.40 & 10.80 & 8.44 & 10.34 & 8.86 & 0.44 \\ 
\end{tabular}}\\
\begin{tabnote}
boot., parametric bootstrap procedure; asym., asymptotic procedure; sd, standard deviation .
\end{tabnote}
\label{tab:m1lvl}
\end{table}

\begin{table}
    \centering
\def~{\hphantom{0}}
\caption{Empirical levels (expressed as percentages) of the test that one variance is null in a linear model with three random effects and one nuisance parameter, for $K=5000$ simulated datasets and $B=500$ bootstrap replicates. The last column gives the maximal standard deviation value obtained in each row.}{
\begin{tabular}{cccccccc}
  \multirow{2}{*}{\shortstack{Level \\ $\alpha$}} & \multicolumn{2}{c}{$\n=20$} & \multicolumn{2}{c}{$\n=30$} & \multicolumn{2}{c}{$\n=40$} & \multirow{2}{*}{\shortstack{max \\ sd}}   \\ 
  & boot & asym & boot & asym & boot & asym & \\
 1\% & 0.82 & 0.66 & 0.72 & 0.58 & 0.90 & 0.62 & 0.13 \\ 
   5\% & 4.46 & 3.54 & 3.96 & 3.28 & 4.14 & 3.08 & 0.29 \\ 
   10\% & 8.88 & 6.78 & 7.52  & 6.34 & 8.40 & 6.98 & 0.40 \\ 
\end{tabular}}\\
\begin{tabnote}
boot., parametric bootstrap procedure; asym., asymptotic procedure; sd, standard deviation.
\end{tabnote}
\label{tab:m2lvl}
\end{table}

\begin{figure}
    \centering
    \begin{subfigure}[b]{0.475\textwidth}
        \includegraphics[scale=0.5]{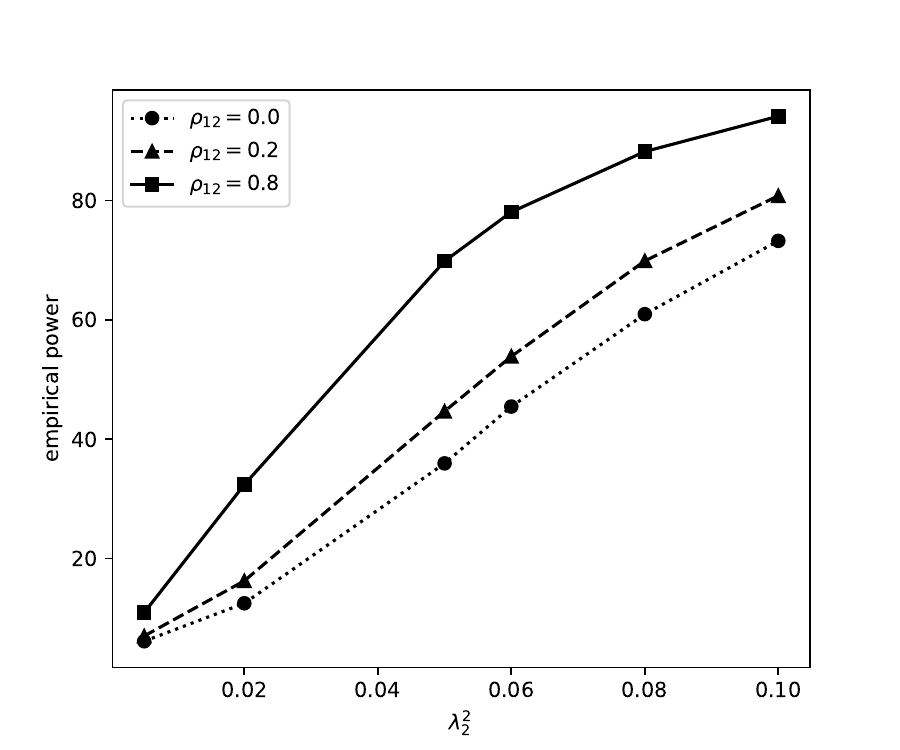} 
        \caption{Model $m_1$}
    \end{subfigure}    
    \begin{subfigure}[b]{0.475\textwidth}
        \includegraphics[scale=0.5]{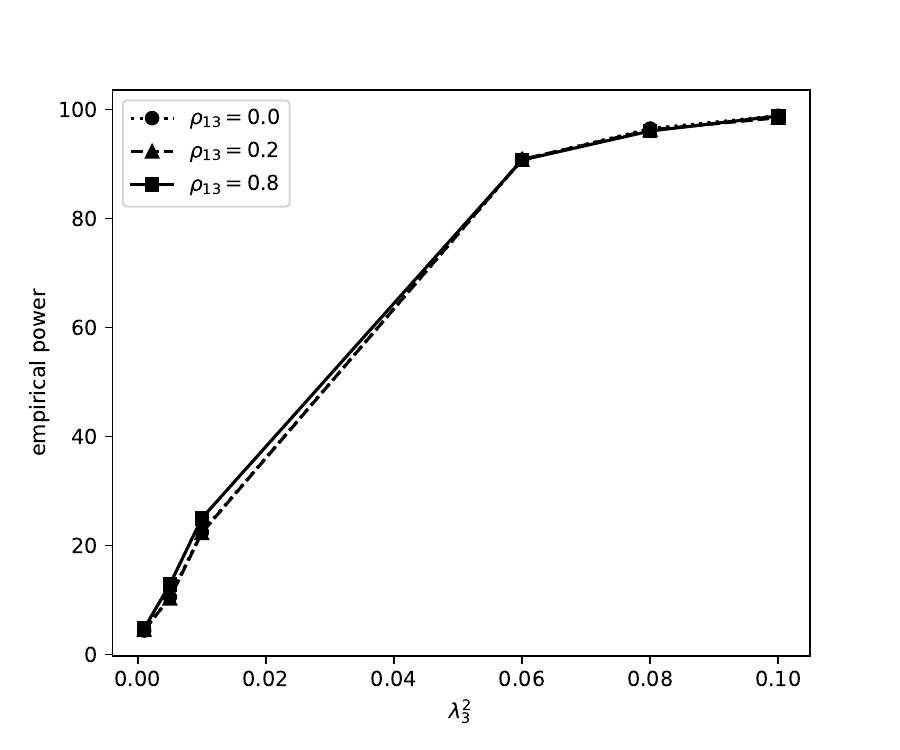}
        \caption{Model $m_2$}
    \end{subfigure}
    \caption{Empirical power of the test that one variance is null in a linear model with (a) two random effects and (b) three random effects, for varying value of the tested variance and of the correlation coefficient, for $K=2500$ simulated datasets and $B=500$ boostrap replicates.}
    \label{fig:power}
\end{figure}

We then study the empirical power of our procedure using models $m_1$ and $m_2$ for $N=30$. To this end, we consider a non diagonal matrix $\Lambda$, introducing a correlation between the components of the scaled random effects $b_i$. We denote by $\rho_{kl}$ the correlation coefficient between the scaled random effects $b_{ik}$ and $b_{il}$.
We then consider increasing values of $\lambda_2$ and $\rho_{12}$ in $m_1$, and increasing values of  $\lambda_3$ and $\rho_{13}$ in $m_2$. 
Results are given in figure \ref{fig:power}. 
As expected, we observe that, for fixed values of the correlation coefficient, the empirical power increases when the true value of the tested variance increases, and that, for fixed values of the variance, the power increases when the correlation coefficient increases. In $m_1$, since $\beta_2=7$, we obtain an empirical power of at least 70\% for a relative standard deviation of 4.5\% (i.e. when $\lambda_2^2 = 0.1$). In $m_2$, since $\beta_3=3$, the empirical power is greater than 12.5\% for a relative standard deviation of 4.7\% (i.e. when $\lambda_3^2 = 0.02$), and above 90\% for a relative standard deviation of 10\% (i.e. when $\lambda_3^2 = 0.1$).

\begin{figure}[ht]
    \centering
    \includegraphics[scale=0.75]{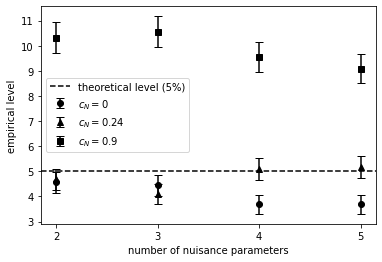}
    \caption{Comparison of the parametric bootstrap procedure and the shrinked parametric bootstrap procedure in $m_3$, using $K=2500$ datasets of size $\n=30$ and $B=300$ bootstrap replicates for varying numbers of nuisance parameters, and different shrinkage parameters $c_\n$.}
    \label{fig:multip}
\end{figure}

We then study the effect of shrinkage on the type I error using model $m_3$. Results are presented in Figure \ref{fig:multip}  for a theoretical level of 5\%. 
We see that the procedure is sensitive to extreme values of $c_\n$. The performances of the shrinked bootstrap procedure are stable as the number of nuisance parameters increases, provided that the shrinkage parameter $c_\n$ is carefully chosen, whereas the performances of the regular bootstrap procedure with no shrinkage are downgraded in this context. Indeed choosing a value of
$c_\n \approx 0.24$, i.e. that shrinks most of the nuisance parameters,
provides good results while choosing $c_\n=0.9$, i.e. of the same order of magnitude as the non-zero variances of the model (here, $\lambda=1$) deteriorates the results. 
On the other hand, neglecting the nuisance parameters, which corresponds to the case $c_\n=0$, also 
as an influence on the results and leads to an empirical level which is smaller than the theoretical one. 

\begin{table}
    \centering

 \caption{Comparison of the bootsrap procedure and the asymptotic procedure in the test in $m_4$, using $K=1000$ datasets of size $\n=40$ and $B=300$ bootstrap replicates.} \label{tab:nonlin}

        \begin{tabular}{ccccc}
 Level $\alpha$ & boot & asym & max sd \\ 
1\% & 0.80 & 0.80 & 0.28 \\ 
  5\% & 5.10 & 3.60 & 0.70 \\ 
  10\% & 10.30 & 7.00 & 0.96 \\ 
\end{tabular}   \\
     \begin{tabnote}
boot., parametric bootstrap procedure; asym., asymptotic procedure; sd, standard deviation.

\end{tabnote}
\end{table}

The previous experiment suggests that shrinking nonzero variances deteriorates the performance of the procedure. However it is difficult to distinguish between nonzero estimations of nuisance parameters and nonzero estimations of small variances. Therefore we carry out a simulation study to investigate  the robustness of the procedure. In particular, we are interested in evaluating through simulation the behavior of the type one error when considering a shrinking parameter of the same order  than the true value of some of the variances. The idea is to mimic a practical shrinking procedure were the threshold would be chosen equal  to  the estimates of the variance parameter. To this end, we consider two experiments using model $m_2$ with  $\beta_0  = (0,7,3)^T$,  $\lambda_{03}=0$ and $\sigma_0^2 = 1.5$. In both cases we test $H_0: \lambda_3=0$ against $H_1:\lambda_3\geq 0$ at the level $5\%$. In the first experiment we set $\lambda_{01} = 1.3$ and consider for $\lambda_{02}$ the different values  $\{0.001, 0.01, 0.01\}$, while in the second experiment we set $\lambda_{02} =1.3$ and consider for $\lambda_{01}$ the different values  $\{0.001, 0.01, 0.01\}$. In the first (respectively second) experiment, $\lambda_2$ (respectively $\lambda_1$) is the supposed nuisance parameter that will be mistakenly shrinked. 
We study the effect of two different shrinkage acting only on the supposed nuisance parameter. In the first case, the shrinkage parameter is chosen of the same order of magnitude as the small untested variance, i.e. we set $c_\n = \lambda_{02}$ (respectively $c_\n = \lambda_{01}$) in the first (respectively second) experiment. 
This leads to a procedure that will often shrink the small variance parameter, even though its true value is strictly nonnegative.  In the second case, we set the bootstrap parameter $\emvB{{\lambda_2}} = 0 $ (respectively $\emvB{{\lambda_1}}=0$) so that the nuisance parameter is always shrinked. Results are presented in Table \ref{tab:robustlambda}. These experiments suggest that shrinking nonzero variances can highly deteriorate the level of the test, especially when the corresponding variance contributes heavily to the total variance of the model.
Indeed, the impact of shrinking a small variance in model $m_2$ is higher when considering $\lambda_2$ rather than $\lambda_1$. This might be due to the fact that the variance associated to one observation is given by $\text{var}(y_{ij}) = \lambda_1^2 + j^2\lambda_2^2 +\sigma^2$ (with $1 \leq j \leq 5 $), so that the part of the total variance that can be attributed to $\lambda_2$ is higher than the part associated to $\lambda_1$.
 However, it also shows that the level can still be preserved when the values of the shrinked variances remain small compared to the total variance given by the model. Overall, this suggests that an expert-based point of view can be adopted for the choice of $c_\n$, guided by the applications. For example, it could be defined as a proportion of the total variance above which the variability of the effect should be taken into account.

\begin{table}
\caption{Empirical levels (expressed as percentages) of the test that one variance is null in a model with three independent random effects $(m_2)$, with one growing nonzero variance parameter shrinked, for $K=2500$ simulated datasets and $B=500$ bootstrap replicates. case(1) corresponds to the case where the shrinkage parameter is equal to the true value of the growing variance, case (2) corresponds to a systematic shrinkage of the growing variance. On the left, the growing variance is $\lambda_2^2$, on the right it is $\lambda_1^2$.} \label{tab:robust_lambda1}
    \centering
    \hfill
    \begin{tabular}{cccc} 
        $\lambda_1^2$ & case (1) & case (2) & max sd  \\
        0.001 & 4.96 & 5.08 & 0.44 \\
        0.01 & 8.92 & 9.04 & 0.57 \\ 
        0.1 & 19.44 & 20.12 & 0.80
    \end{tabular}
    \hfill
    \begin{tabular}{cccc} 
        $\lambda_2^2$ & case (1) & case (2) & max sd\\
        0.001 & 4.16 & 4.84 &  0.43 \\
        0.01 & 5.4 & 5.48 & 0.46 \\ 
        0.1 & 5.24 & 5.76 &  0.47
    \end{tabular}
    \hfill\null
    \label{tab:robustlambda}\\
    \begin{tabnote}
    \centering{sd, standard deviation}
    \end{tabnote}
\end{table}

\subsection{Real data application}
We then apply our procedure to a study of white-browed coucal growth rates, available as a Dryad package \citep{goymann2016sex}. 
We use the logistic growth model defined in \eqref{eq:logisticmodel} to describe the evolution of the body mass of the nestlings as a function of their age. More precisely, if we denote by $\obs_{ij}$ the body mass of nestling $i$ at age $\temps_j$ for $i=1,...,292$, $j=1,...,\n_i$, we have that $\beta_1+\lambda_1\xi_{i1}$ is the asymptotic nestling body mass, $\beta_2+\lambda_2\xi_{i2}$ is the age (in days) at which nestling $i$ reaches half its asymptotic body mass and $\beta_3+\lambda_3\xi_{i3}$ is the growth rate of nestling $i$. When fitting the complete model the  estimated scaling matrix  is $\hat{\Lambda} = \diag(\sqrt{212.34}, \sqrt{0.89}, \sqrt{0.02})$, which motivates the test that $\lambda_2$ and $\lambda_3$ are null. 

In order to test for the presence of randomness in the inflexion point and in the growth rate, we proceed sequentially. First, we test if the variances of both the inflexion point and the growth rate are null, i.e. we consider the test $T_1: H_0: \lambda_2 = 0, \lambda_3 = 0 $ against $H_1: \lambda_2 \geq 0, \lambda_3\geq 0$. If the null hypothesis in $T_1$ is rejected, we perform two univariate tests, one for each variance tested in $T_1$. More precisely, we consider the tests $T_2: H_0: \lambda_3 = 0 $ against $H_1 : \lambda_3\geq 0$ and $T_3: H_0: \lambda_2 = 0 $ against $H_1 : \lambda_2\geq 0$. If the null hypothesis is rejected in $T_1$, it means that at least one of the two tested variances is nonzero, thus we can then consider the univariate tests $T_2$ and $T_3$, where $\lambda_2$ and $\lambda_3$ might be nuisance parameters. For each test, we consider two procedures, one where the estimate of the potential nuisance parameter ($\lambda_2$ in $T_2$ and $\lambda_3$ in $T_3$) is shrinked toward zero, and another one without shrinkage. This choice enables to consider the two possible cases : $\lambda_2>c_\n$ and $\lambda_2<c_\n$ for $T_2$ and $\lambda_3>c_\n$ and $\lambda_3<c_\n$ for $T_3$. In practice the practitioner can choose the threshold according to his own level of significance of variability desired. For instance by saying that under X\% of relative variability of a parameter, we consider it as a fixed effect. Table \ref{donnees reelles} compiles the results of the three tests. 


\begin{table}[h]
    \centering
\def~{\hphantom{0}}
\caption{Comparison of the $p$--value (in \%) among the three tests $T1$, $T2$ and $T3$, using $B=1000$ bootstrap replicates.}{
\begin{tabular}{cccccc}
 Test & T1 & \multicolumn{2}{c}{T2} & \multicolumn{2}{c}{T3}  \\ 
$\textsc{lrt}$ & $302.5$ & \multicolumn{2}{c}{1.6}  &\multicolumn{2}{c}{278.2} \\
 procedure & no shrink & shrink & no shrink &   shrink & no shrink\\
 $p$--value & 0 & 8.9 & 7.9 & 0& 0\\ 
\end{tabular}}\\
\begin{tabnote}
shrink., shrinked parametric bootstrap procedure; no shrink., regular parametric bootstrap procedure without shrinkage
\end{tabnote}
\label{donnees reelles}
\end{table}

We can see that the procedures lead to different $p$-values, and can thus, in practice, lead to different conclusions with respect to the null hypothesis depending on the type I error considered.  

\section{Discussion}
This work can lead to several future developments, both from a theoretical and a practical point of view. In particular, the choice of the shrinking parameter $c_\n$ is of interest for practitioners that would want to apply our procedure. We showed that in some cases it can have a large impact on the estimated level, especially when it leads to the shrinkage of variances that amount for a large part of the total variance. This suggests that the choice of this tuning parameter could be tackled from an expert-based point of view, based on the expected variability of each parameter. For example, $c_\n$ could be defined as a threshold under which the variability of a random effect is not relevant for the task of interest, as long as it does not account for a significant part of the total variance. However, this compromise is model-dependent and should be considered according to each specific application.  From a methodological point of view, it would be interesting to propose an automated procedure, following for example the idea of \cite{bickel2008choice}.  

In a model building approach, our procedure presents the advantage of dealing with nuisance parameters, therefore sequential tests can be performed as we did in the real data applications. This idea is promising to select the exact number of random effects to consider, however such sequential tests present other issues which require further development to be addressed carefully  such as multiple testing and  post selection concerns that are beyond  the scope of this paper. 

From a computational point of view, our algorithm can be used indifferently for linear and nonlinear mixed effects models, however the computation time can be prohibitive in the latter case, especially for complex models. It would be interesting to find a criterion to optimize the choice of the bootstrap sample size.  Another issue when dealing with complex models is the computation of the likelihood ratio test statistic. Indeed, it involves a ratio of likelihoods which are usually estimated separately using Monte Carlo approaches, leading to a biased estimate. This point is crucial since this quantity is calculated at each iteration of the bootstrap procedure.

\section{Acknowledgment}

This work was funded by the Stat4Plant project ANR-20-CE45-0012.

\begin{appendices}
    
\section{Supplementary material}

We recall that we consider the following nonlinear mixed effects model, for any $i=1,...,\n$ and any $j=1,..,\J$:

\begin{equation}\label{eq:model2}
    \left\{ 
    \begin{array}{ll}
          \obs_{ij} &= g(\cov_{ij},\beta,\Lambda\xi_i)+\varepsilon_{ij}\quad \varepsilon_{ij} \sim \mathcal{\n}(0,\sigma^2) \\
         \xi_i & \sim\mathcal{\n}\left(0,\Id{\df} \right) \\
    \end{array} \right.
\end{equation} 

We also recall that the log--likelihood of the model given the data $\obs_{1:\n}$ writes: 
\begin{equation}\label{eq:likelihood3}
    l(\theta ; \obs_{1:\n}) =\log\{L_\theta(\obs_{1:\n})\}= \log\{\prod_{i=1}^{\n}\f\}= \sum_{i=1}^{\n}\log\{\int \fc\theta \pi(\xi_i)d\xi_i\}.
\end{equation} 

Finally we write
\begin{equation}\label{eq:emv}
    \emv\theta = \arg \underset{\theta\in\Theta}{\sup}\quad l(\theta;\obs_{1:\n}).
\end{equation}

\subsection{Different parametrizations for mixed effects models }\label{sec:diffparam}

The commonly used parametrization for nonlinear mixed effects models  (\cite{pinheiro2006mixed} page 306) is for $i=1,...,\n$, $j=1,...,\J$ : 
\begin{equation}
    \left\{ 
    \begin{array}{ll}
          \obs_{ij}  &= g(v_{ij},\phi_i) + \varepsilon_{ij}\quad \varepsilon_{ij} \sim \mathcal{\n}(0,\sigma^2) \\
         \phi_i &= A_{ij}\beta+B_{ij}b_i \quad b_i\sim\mathcal{\n}\left(0,\Gamma \right) \\
    \end{array} \right.
\end{equation}

where $v_{ij}$, $A_{ij}$ and $B_{ij}$ are known covariates. With this definition, the log--likelihood is defined as follows : 

\begin{equation}\label{eq:likelihood2}
    l(\theta ; \obs_{1:\n}) =\log\{L_\theta(\obs_{1:\n})\}= \log\{\prod_{i=1}^{\n}\f\}= \sum_{i=1}^{\n}\log\{\int f_i(\obs_i;b_i,\theta)\pi(b_i;0,\Gamma)db_i\},
\end{equation}

where $f_i(\obs_i;b_i,\theta)$ is the density of the conditional distribution of $\obs_i$ given $b_i$, and $\pi(b_i;0,\Gamma)$ is the density of the $p$-dimensional centered Gaussian distribution with covariance $\Gamma$. 

The change of variable $b_i = \Lambda\xi_i$ is a $\mathcal{C}^1$ diffeomorphism if and only if the diagonal coefficients of $\Lambda$ are strictly nonnegative. Therefore in our setting where this condition is not verified these two parametrizations are not equivalent.  

In particular taking $\Lambda=0$ in \eqref{eq:likelihood} is equivalent to considering a fixed-effects nonlinear model, while in \eqref{eq:likelihood2}, taking $\Gamma\rightarrow 0$ makes $l(\theta ; \obs_{1:\n})\rightarrow -\infty$.

\subsection{Proof of proposition \ref{lem:singularity}}\label{app:prop1proof}

To prove the proposition we only have to prove that if the $m$th column of $\Lambda$ is $0$, then the odds orders of the partial derivatives of the log-likelihood with respect to the elements of this column are null regardless of the values of the $\obs_{1:\n}$. We have, for all $n = 1,..,\df$:

\begin{equation*}
    \frac{\partial \f}{\partial [\Lambda]_{nm}}|_{\theta=\theta_0}  = \int_{\xi_1}\cdot\cdot\cdot\int_{\xi_p} \frac{\partial f_i(\obs_i;\xi,\theta_0)}{\partial [\Lambda]_{nm}}\pi(\xi_1)d\xi_1...\pi(\xi_\df)d\xi_\df  
\end{equation*}

Given the definition of model \ref{eq:model} :

\[f_i(\obs_i;\xi,\theta) = (2\pi\sigma^2)^{-\frac{\J}{2}}\\exp\left[-\frac{\sum_{j=1}^{\J}\left\{\obs_{ij}-g(\cov_{ij},\beta,\Lambda\xi)\right\}^2}{2\sigma^2}\right]\]

therefore,

\begin{align*}
    \frac{\partial f_i(\obs_i;\xi,\theta)}{\partial [\Lambda]_{nm}} &\propto \frac{\partial \Lambda\xi}{\partial [\Lambda]_{mn}}\nabla_{\Lambda\xi}\exp\left[-\frac{\sum_{j=1}^{\J}\left\{\obs_{ij}-g(\cov_{ij},\beta,\Lambda\xi)\right\}^2}{2\sigma^2}\right]\\
    &\propto \xi_m\nabla_{\Lambda\xi}\exp\left[-\frac{\sum_{j=1}^{\J}\left\{\obs_{ij}-g(\cov_{ij},\beta,\Lambda\xi)\right\}^2}{2\sigma^2}\right]
\end{align*}

Evaluated at $\theta=\theta_0$ the last term (the gradient) no longer depends on $\xi_m$ as the $m$th column of $\Lambda$ is null. 

\begin{align*}
    \frac{\partial \f}{\partial [\Lambda]_{nm}}|_{\theta=\theta_0} \propto &\int_{\xi_m} \xi_m \pi(\xi_m)d\xi_m \\
    &\times\int\nabla_{\Lambda\xi}\exp\left[-\frac{\sum_{j=1}^{\J}\left\{\obs_{ij}-g(\cov_{ij},\beta,\Lambda\xi)\right\}^2}{\sigma^2}\right]\mid_{\theta=\theta_0}\prod_{l\neq m}\pi(\xi_l)d\xi_l
\end{align*}

which is equal to $0$ as the first term of the right hand side equation is expectation of a standard gaussian distribution. We can apply the same reasoning to every odds order derivatives.  

\subsection{Proof of proposition \ref{prop:consist mle}}

Due to assumption \eqref{ass:regularity1} we have that :

\begin{equation}\label{identifiability}    
 \underset{\theta\in\Theta}{\sup}\E \left\{l(\theta;\obs_1)\right \} < \E\left\{l(\theta_0;\obs_1)\right\}
\end{equation}

which comes from the identifiability of the model and the positivity of the Kullback-Leibler divergence. Assumption \eqref{ass:regularity 0-4}$i)$ enables to apply the uniform law  of large number to the log--likelihood. Then the result follows from arguments as in \cite{andrews1993tests} lemma A.1.

\subsection{Proof of proposition \ref{prop:consistboot}}

To prove the consistency of the bootstrap maximum likelihood estimator, we will use the same reasoning, as in the proof of proposition \ref{prop:consist mle}. The sketch of the proof is similar to the one of \cite{cavaliere2020bootstrap}. 

We first want to show that :

\[\underset{\theta\in\Theta}{\sup}| \frac{1}{\n}l(\theta;y^*_{1:\n})-\E\left\{l(\theta;\obs_1)\right\}| =o_{p^*}(1) \]

First of all we have that : 

\begin{align*}
\underset{\theta\in\Theta}{\sup}| \frac{1}{\n}l(\theta;y^*_{1:\n})-\E\left\{l(\theta;\obs_1)\right\}| &\leq \underset{\theta\in\Theta}{\sup} A_\n^*(\theta) + \underset{\theta\in\Theta}{\sup} A_\n(\theta)
\end{align*}

where : 

\begin{align*}
    A_\n^*(\theta) & = | \frac{1}{\n}l(\theta;y^*_{1:\n})- \E^*\left\{l(\theta;\obs_1^*)\right\}| \\
    A_\n(\theta) &= |  \E^*\left\{l(\theta;\obs_1^*)\right\}-\E\left\{l(\theta;\obs_1)\right\}|
\end{align*}

We now want to apply the uniform law of large numbers to $A_\n(\theta)$, and it's bootstrap version to $A_\n^*(\theta)$. Therefore we shall show that both term converges toward $0$ and that they are lipshitz. \\
We first consider $A_\n(\theta)$ for a given $\theta\in\Theta$. 

\begin{align*}
     \E^*\left\{l(\theta;\obs_1^*)\right\} &= \E\{l(\theta;\obs_1^*)|\obs_{1:\n}\} \\
    &= \int l(\theta;\obs_1)f(\obs_1; \emvB{\theta})d\obs_1
\end{align*}

Using assumption \eqref{ass:regularity1}, we can state that there exist $\theta^+$ between $\theta_0$ and $\emvB{\theta}$ such that, 

\begin{align*}
|f(\obs_1; \emvB{\theta})-f(\obs_1; \theta_0)| &\leq \|\theta_0-\emvB{\theta}\| \|\nabla_\theta f(\obs_1; \theta^+)\| \\
&\leq  \|\theta_0-\emvB{\theta}\| \|\nabla_\theta \log\{f(\obs_1; \theta^+)\}\|f(\obs_1; \theta^+)
\end{align*}

therefore, 

\begin{align*}
    A_\n(\theta) &\leq \int |l(\theta;\obs_1)|||f(\obs_1; \emvB{\theta}) - f(\obs_1;\theta_0)|d\obs_1 \\
    &\leq \int |l(\theta;\obs_1)|\|\theta_0-\emvB{\theta}\| \|\nabla_\theta \log\{f(\obs_1; \theta^+)\}\|f(\obs_1; \theta^+)d\obs_1\\
    &\leq \|\theta_0-\emvB{\theta}\| \int |l(\theta;\obs_1)|\|\nabla_\theta l(\theta^+;\obs_1)\}\|f(\obs_1; \theta^+)d\obs_1
\end{align*}

Due to assumption \eqref{ass:regularity 0-4}$(i)$--$(ii)$, $|l(\theta;\obs_1)|\|\nabla_\theta l(\theta^+;\obs_1)\}\|$ is integrable with respect to the density $f(\obs_1; \theta^+)$. And finally using the elementary inequality , 

\begin{equation}\label{eq:outil1}
\frac{1}{2}(a^2+b^2)\geq |ab|,\quad\forall a,b\in\bbr
\end{equation}

we can state that 

\begin{align*}
    A_\n(\theta) &\leq \frac{1}{2} \|\theta_0-\emvB{\theta}\| \underset{\theta^+\in \Theta}{\sup}\int \underset{\Theta\in\Theta}{\sup}|l(\Theta;\obs_1)|^2+\underset{\theta_2\in\Theta}{\sup}\|\nabla_\theta l(\theta_2;\obs_1)\}\|^2f(\obs_1; \theta^+)d\obs_1
\end{align*}

Finally thanks to assumption \eqref{prop:consist mle}, as $\n \rightarrow +\infty$, it holds in probability that :

\[A_\n(\theta)\rightarrow 0\]

We now consider :

\[A_\n^*(\theta) = |\frac{1}{\n}\sum_{i=1}^\n l(\theta;y_i^*)- \E^*\left\{l(\theta;\obs_1^*)\right\}|\]

This quantity is a sum of conditionally independent and centered random variables. We can't directly apply a law of large number as the parameter $\emvB{\theta}$ and the index of the sum depends both on $\n$. 

For every real nonnegative number $t$, it holds almost surely that : 

\[\pr^*(A_\n^*>t)\leq \pr^*\{\frac{1}{\n}\sum_i| l(y_i^*;\theta)- \E^*\left\{l(\theta;\obs_1^*)\right\}|>t\}\leq \frac{\underset{\theta'\in\Theta}{\sup} \E_{\theta'}\{\underset{\theta\in\Theta}{\sup}|\lz|^2\}}{ \n t^2}\]

by applying first triangular inequality and then Chebychev inequality, using assumption \eqref{ass:regularity 0-4}$i)$. And finally $A_\n^*(\theta)\rightarrow 0$ in probability, as $\n\rightarrow+\infty$,  which concludes the pointwise convergence. And we note that this result holds uniformly over $\Theta$ so:

\[\underset{\theta\in\Theta}{\sup} | \frac{1}{\n}l(\theta;y^*_{1:\n})-\E\left\{l(\theta;\obs_1)\right\}| =o_{p^*}(1) \]

Let now use this result to show that the bootstrap maximum likelihood estimator is consistent. 

 Let $\varepsilon>0$, using equation \eqref{identifiability}, there exists $\delta>0$ such that: 
 
\[\underset{\|\theta-\theta_0\|>\varepsilon}{\inf} \E\{l(\theta_0;\obs_1)) - \E(l(\theta;\obs_1)\}\geq \delta\] 

Let us introduce $\theta_{mle}^B= \arg \underset{\theta\in\Theta}{\max}\quad \lnb$.

By writing $V_\varepsilon = \{\theta\in\Theta:\|\theta-\theta_0\|>\varepsilon\}$, we have that : 

\begin{align*}
    \pr^*(\theta_{mle}^B\in V_\varepsilon ) &\leq \pr^*\left[\E\{l(\theta_0;\obs_1)\} - \E\{l(\theta_{mle}^B;\obs_1^*)\}\geq \delta\right]\\
    & = \pr^*\left[\E\left\{l(\theta_0;\obs_1)\right\} - \frac{1}{\n}l(\theta_{mle}^B;\obs_{1:\n}^*) + \frac{1}{\n}l(\theta_{mle}^B;\obs_{1:\n}^*)- \E\left\{l(\theta_{mle}^B;\obs_1^*)\right\}\geq \delta\right] \\
    & \leq \pr^*\left[\E\{l(\theta_0;\obs_1)\}  - \frac{1}{\n}l(\theta_0;y^*_{1:\n}) + \frac{1}{\n}l(\theta_{mle}^B;\obs_{1:\n}^*)- \E\{l(\theta_{mle}^B;\obs_1^*)\}\geq \delta\right] \\
    & \leq \pr^*\left\{2 \underset{\theta\in\Theta}{\sup} | \frac{1}{\n}l(\theta;y^*_{1:\n})-\E\left\{l(\theta;\obs_1)\right\}|\geq \delta\right\} \\
    & \leq o_{p}(1)
\end{align*}

Which concludes the proof that $\theta_{mle}^B = \theta_0 + o_{p^*}(1)$. The exact same proof still holds for the restricted bootstrap maximum likelihood estimator by replacing $\Theta$ by $\Theta_0$.

\subsection{Quadratic approximation of the log--likelihood}

To derive the asymptotic distribution of $LRT_\n$, we expand the log-likelihood around $\theta_0$ (see \cite{andrews1999estimation} theorem 6). Under assumption \eqref{ass:regularity1}, following the lines of \cite{testinghiroyuki2012}, we can write: 

\begin{align*}
    l(\theta;\obs_{1:\n}) - &l(\theta_0;\obs_{1:\n}) = \diff{\psi}^T\dpsi + \frac{1}{2}\diff{\psi}^T\dpsii\diff{\psi} \\
   & + (1/2)\diid{i}{j} + (3/3!)\diff{\psi}^T\sum_{i,j=1,...,d_\delta}\delta_{i} \delta_{j}\frac{\partial^3 l(\theta_0;\obs_{1:\n})}{\partial\delta_{i}\partial\delta_{j}\partial\psi}\\
    & + (1/2)\dii{i}{j} + (3/3!)\diff{\psi}^T\sum_{i,j=1,...,d_\lambda}\lambda_{i} \lambda_{j}\frac{\partial^3 l(\theta_0;\obs_{1:\n})}{\partial\lambda_{i}\partial\lambda_{j}\partial\psi}\\
       & + (6/4!)\sum_{i,j=1,...,d_\delta}\sum_{k,l=1,...,d_\lambda} \delta_i\delta_j\lambda_k\lambda_l \frac{\partial^4 l(\theta_0;\obs_{1:\n})}{\partial\delta_i\partial\delta_j\partial\lambda_k\partial\lambda_l} \\
       & + (1/4!)\diiiid{i}{j}{k}{l} \\
    & + (1/4!)\diiii{i}{j}{k}{l} + R_\n(\theta)
    \end{align*}

   With $R_\n(\theta)$ being the rest in the Taylor expansion . 
   We define for all integers $i,j$ $c_{ij} = \frac{1}{2}$ if $i=j$ and $1$ otherwise. We also define $\mathcal{I}_\lambda= \{11,22, ..,d_\lambda -1, d_\lambda\}$ and $\mathcal{I}_\delta= \{11,22, ..,d_\delta -1, d_\delta\}$ that respectively index $v(\lambda)=(\lambda_i\lambda_j)_{ij\in\mathcal{I}_{\lambda}}$ and $v(\delta)=(\delta_i\delta_j)_{ij\in\mathcal{I}_{\delta}}$. For $ij\in \mathcal{I}_\lambda$ (respectively $\mathcal{I}_\delta$), we write $\frac{\partial^2l(\theta;\obs_{1:\n})}{\partial v(\lambda)_{ij}} =\frac{\partial^2l(\theta;\obs_{1:\n})}{\partial \lambda_i\lambda_j} $ (the same with $\delta$). With these notations we have that: 
 
\begin{align*}
    l(\theta;\obs_{1:\n}) - l(\theta_0;\obs_{1:\n}) = &\diff{\psi}^T\dpsi + \frac{1}{2}\diff{\psi}^T\dpsii\diff{\psi} \\
    & + \sum_{i\in\mathcal{I}_\lambda}v(\lambda)_i c_i \frac{\partial^2l(\theta;\obs_{1:\n})}{\partial v(\lambda)_{i}} + \diff{\psi}^T\sum_{i\in\mathcal{I}_\lambda}v(\lambda)_{i}c_i \frac{\partial^3 l(\theta_0;\obs_{1:\n})}{\partial v(\lambda)_{i}\partial\psi}\\
    & +  \sum_{i\in\mathcal{I}_\delta}v(\delta)_i c_i \frac{\partial^2l(\theta;\obs_{1:\n})}{\partial v(\delta)_{i}} + \diff{\psi}^T\sum_{i\in\mathcal{I}_\delta}v(\delta)_{i}c_i \frac{\partial^3 l(\theta_0;\obs_{1:\n})}{\partial v(\delta)_{i}\partial\psi}\\
    & + 4\times (6/4!)\sum_{i\in\mathcal{I}_\lambda, j\in \mathcal{I}_\delta}c_ic_jv(\lambda)_iv(\delta)_j\frac{\partial^4 l(\theta_0;\obs_{1:\n})}{\partial v(\lambda)_i\partial v(\delta)_j} \\
    & + (4/4!)\sum_{i\in\mathcal{I}_\delta, j\in \mathcal{I}_\delta} c_ic_jv(\delta)_iv(\delta)_j \frac{\partial^4l(\theta_0;\obs_{1:\n})}{\partial v(\delta)_i\partial v(\delta)_j} \\
    & + (4/4!)\sum_{i\in\mathcal{I}_\lambda, j\in \mathcal{I}_\lambda} c_ic_jv(\lambda)_iv(\lambda)_j \frac{\partial^4l(\theta_0;\obs_{1:\n})}{\partial v(\lambda)_i\partial v(\lambda)_j}  + R_\n(\theta)\\
    \end{align*}

We define the reparametrization of $\theta$ : $\phi(\theta) = (\psi,v(\delta),v(\lambda))$.

 By writing $ \tilde{\nabla}_{v(\lambda)}l(\theta;\obs_{1:\n}) = \left(c_{i}\frac{\partial^2l(\theta;\obs_{1:\n})}{\partial v(\lambda)_{i}}\right)_{i\in\mathcal{I_\lambda}}$ (similarly for $\tilde{\nabla}_{v(\delta)}l(\theta;\obs_{1:\n})$) and $ \tilde{\nabla}^2_{v(\lambda)}l(\theta;\obs_{1:\n}) = (c_{i}c_{j}\frac{\partial^4l(\theta;\obs_{1:\n})}{\partial v(\lambda)_{i}\partial v(\lambda)_{j}})_{i,j=1,...,d_\lambda}$ (similarly for $\tilde{\nabla}^2_{v(\delta)}l(\theta;\obs_{1:\n})$ ), we also define :
 
\begin{equation*}
    \tilde{S}_\n(\theta_0) = \sqrt{\n}^{-1}\left(\begin{tabular}{ll}
            $\nabla_{\psi}l(\theta_0;\obs_{1:\n})^T$,
          $\tilde{\nabla}_{v(\delta)}l(\theta_0;\obs_{1:\n})^T$,
          $\tilde{\nabla}_{v(\lambda)}l(\theta_0;\obs_{1:\n})^T$
    \end{tabular}\right)^T
\end{equation*}

\begin{equation}\label{eq:IN}
    \tilde{I}_\n(\theta_0) = \begin{pmatrix}
         I_{\n,\psi}(\theta_0) & I_{\n,\psi,v(\delta)}(\theta_0) & I_{\n,\psi,v(\lambda)}(\theta_0)  \\
         I_{\n,\psi,v(\delta)}(\theta_0)^T & I_{\n,v(\delta)}(\theta_0)& I_{\n,v(\delta),v(\lambda)}(\theta_0)  \\ 
         I_{\n,\psi,v(\lambda)}(\theta_0)^T & I_{\n,v(\delta),v(\lambda)}(\theta_0)^T  & I_{\n,v(\lambda)}(\theta_0)
    \end{pmatrix}
\end{equation}

Where $I_{\n,\psi}(\theta_0) = -\frac{1}{\n}\nabla^2_{\psi}l(\theta_0;\obs_{1:\n})$, $I_{\n,\psi,v(\lambda)}(\theta_0) =  \left(-\frac{c_{i}}{\n}\frac{\partial^3l(\theta_0;\obs_{1:\n})}{\partial \psi\partial v(\lambda)_{i}}\right)_{.,i\in\mathcal{I}_\lambda}$, $I_{\n,\psi,v(\delta)}(\theta_0) =  \left(-\frac{c_{i}}{\n}\frac{\partial^3l(\theta_0;\obs_{1:\n})}{\partial \psi \partial v(\delta)_{i}}\right)_{.,i\in\mathcal{I}_\delta}$, $ I_{\n,v(\lambda)}(\theta_0)=  -\frac{1}{3\n} \tilde{\nabla}^2_{v(\lambda)}l(\theta_0;\obs_{1:\n})$, 
$I_{\n,v(\delta)}(\theta_0)=  -\frac{1}{3\n} \tilde{\nabla}^2_{v(\delta)}l(\theta_0;\obs_{1:\n})$, $I_{\n,v(\lambda), v(\delta)}(\theta_0) =(-\frac{1}{\n}c_ic_j\frac{\partial^4 l(\theta_0;\obs_{1:\n})}{\partial v(\lambda)_i\partial v(\delta)_j})_{i\in\mathcal{I}_\lambda, j\in\mathcal{I}_\delta}$.

Where the notation $(a_i)_{.,i\in \mathcal{I}}$ stands for a $dim(a_i)\times card(\mathcal{I})$ matrix whose columns are $a_i$ for $i\in\mathcal{I}$.

With these new notations we obtain the following quadratic approximation of the log--likelihood :

\begin{equation}\label{dev}
\begin{split}
    l(\theta;\obs_{1:\n}) - l(\theta_0;\obs_{1:\n})  = \sqrt{\n}(&\phi(\theta)-\phi(\theta_0)) ^T \tilde{S}_\n(\theta_0) \\ 
    & -\frac{1}{2}\sqrt{\n}(\phi(\theta)-\phi(\theta_0)) ^T\tilde{I}_\n(\theta_0)\sqrt{\n}(\phi(\theta)-\phi(\theta_0))  + R_\n(\theta)
\end{split}
\end{equation}

\subsection{Asymptotic distribution of the likelihood ratio test statistic}\label{sec:Alrtasym}

We start from the expansion of the log--likelihood \eqref{dev} derived in the last section, that we rewrite:

\begin{equation*}
    \begin{split}
            l(\theta;\obs_{1:\n}) - l(\theta_0;\obs_{1:\n}) = \frac{1}{2}&Z_\n(\theta_0)^T \tilde{I}_\n(\theta_0)Z_\n(\theta_0)\\ 
    &-  \frac{1}{2}\left(t_\n(\theta) - Z_\n(\theta_0)\right)^T\tilde{I}_\n(\theta_0)\left(t_\n(\theta) - Z_\n(\theta_0)\right)+ R_\n(\theta)
    \end{split}
\end{equation*}

where  $t_\n(\theta) = \sqrt{\n}(\phi(\theta) - \phi(\theta_0))$ and $Z_\n(\theta_0)=\tilde{I}_\n(\theta_0)^{-1}\tilde{S}_\n(\theta_0)$

\begin{remark}
     We consider the quantity $\tilde{I}_\n(\theta_0)^{-1}$ which implies that $\tilde{I}_\n(\theta_0)$ is non-singular which may not be always true. However under assumption \eqref{ass:Itilde}, the probability that $\tilde{I}_\n(\theta_0)$ is non-singular tends to 1 as $\n\rightarrow+\infty$. 
\end{remark}

The set a feasible values for $t_\n(\theta)$ is:

\begin{equation}\label{eq:cone}
t_\n(\Theta)= \{\sqrt{\n}\left(\Theta_\psi-\psi_0\right)\}\times\{\sqrt{\n}\left(v(\Theta_\lambda)-v(\lambda_0)\right)\}\times\{\left(v(\Theta_\delta)-v(\delta_0)\right)\}    
\end{equation}

$t_\n(\Theta)$  does not depend on $\n$ ( it is a cartesian product whose terms are whether $\mathbb{R}$ or $[0,+\infty[$),  and it is locally approximated by a cone (in fact it is a cone), we write it $\mathcal{C}(\Theta)$. Therefore if we prove that $R_\n(\theta)$ is $o_p(1)$ when evaluated at the maximum likelihood estimator,   we can apply the result from \cite{andrews1999estimation}. For more details see \citet[Theorem 3]{andrews1999estimation} and paragraph 4.3.

In order to apply the theory of Andrews, one shall prove that $\tilde{I}_\n(\theta_0)$ converges in probability toward a nonnegative matrix $\tilde{I}(\theta_0)$, and that $\tilde{S}_\n(\theta_0)$ converges weakly to a random variable $U(\theta_0)$.

1)\quad Let $d_{\tilde{I}} \in\mathbb{N}$ such that  $\tilde{I}_\n(\theta_0)\in\mathbb{R}^{d_{\tilde{I}} \times d_{\tilde{I}} }$, let $1\leq m,n\leq d_{\tilde{I}} $. We can write: 

\[\left[\tilde{I}_\n(\theta_0)\right]_{m,n} = \frac{1}{\n}\sum_{i=1}^{\n}h_{m,n}^{(i)}(\theta_0)\]

where $h_{m,n}^{(i)}(\theta_0)$ is of the form:
\[h_{m,n}^{(i)}(\theta_0) = c_{m,n}\frac{\partial^k\log f(\obs_i;\theta)}{\partial \theta_{i_1}...\partial\theta_{i_k}}|_{\theta= \theta_0}\]
with $c_{m,n}\in\mathbb{R}$, $k\in\{2,4\}$, $1\leq i_1,...,i_k\leq d_\lambda +d_\psi+d_\delta$.

With assumption \eqref{ass:regularity 0-4} we can apply the law of large numbers to this empirical mean and therefore it holds in probability that:
\[\tilde{I}_\n(\theta_0) \underset{\n \rightarrow +\infty}{\longrightarrow} \left[\E\{h_{m,n}^{(1)}(\theta_0)\}\right]_{1\leq m,n\leq q_I} = \tilde{I}(\theta_0)\]

2) We now consider $\tilde{S}_\n(\theta_0)$.

First the score is centered,

   \[\E\left[\nabla_\psi \log f(\cov_i;\theta_0)\right] = 0\]

then, due to proposition \eqref{lem:singularity}, $\forall m,n = 1,...,q_\lambda$,
   
    \[\E\left[\frac{\partial^2\log f(\obs_i;\theta_0) }{\partial \lambda_m\partial\lambda_n}\right] =-\E\left[\frac{\partial\log f(\obs_i;\theta_0)}{\partial\lambda_m}\frac{\partial\log f(\obs_i;\theta_0)}{\partial\lambda_n}\right] = 0 \]

We apply the central limit theorem  to $\tilde{S}_\n(\theta_0)$ which is a sum of independent and identically distributed  centered random variables with finite variances, and therefore $\tilde{S}_\n(\theta_0)$ converges weakly to a random variable $U(\theta_0)$ 

By doing the exact same quadratic approximation and development but considering the parameter space $\Theta_0$, combining \eqref{eq:cone}--\eqref{eq:reste}, we use \cite{andrews1999estimation} to obtain : 

\begin{equation}\label{eq:lrtasym}
  LRT_\infty = \underset{t\in\mathcal{C}(\Theta_0)}{\inf}\|t-\tilde{I}(\theta_0)^{-1}U(\theta_0)\|_{\tilde{I}(\theta_0)}-\underset{t\in\mathcal{C}(\Theta)}{\inf}\|t-\tilde{I}(\theta_0)^{-1}U(\theta_0))\|_{\tilde{I}(\theta_0)}  
\end{equation}

Which leads to the expression of $LRT_\infty$. 

\begin{remark}\label{rem:nonsing IN}
    On the matrix $\tilde{I}(\theta_0)$, it is obviously symmetric as the limit of \eqref{eq:IN}.It is also positive semi-definite because one can show that $\tilde{I}(\theta_0)$ is the asymptotic variance of $\tilde{S}_\n(\theta_0)$. The proof can be found in a simpler case in the proof of proposition 2 (b) of \cite{testinghiroyuki2012}. The proof is based on the Fisher's identity and the fact that the odd derivatives with respect to $\lambda$ and $\delta$ are zero. 
\end{remark}

\subsection{Proof of proposition \ref{prop:speed}}

To obtain an explicit form for $R_\n(\theta)$ we use the multivariate  version of Taylor-Lagrange formula, which is for instance defined in \cite{andrews1999estimation} Theorem 6. 

This way we have that $R_\n(\theta)$ is a sum of higher order derivatives with respect to $\psi$ and the fourth crossed derivatives with respect to $\lambda$. Given assumption \eqref{ass:regularity 0-4} all the derivatives of the log-likelihood are $\mathcal{O}_p(\n)$, using Cauchy-Schwartz inequality and the fact that  $\|t_\n(\theta)\|^2 =\n\left( \|\psi-\psi_0\|^2+\|v(\delta)\|^2 +\|v(\lambda)\|^2\right)= \n\left(\|\psi-\psi_0\|^2+ \|\lambda\|^4+ \|\delta\|^4\right)\mathcal{O}(1)$ we have that: 

\begin{align*}
    |R_\n(\theta)|\leq \mathcal{O}_p(\n)(  \|\psi-\psi_0\|^3 &+ \|\psi-\psi_0\|^4 +\|\psi-\psi_0\|^2 \|\lambda\|^2+\|\psi-\psi_0\|^2 \|\delta\|^2)\\
    &+ \|\delta\|^4  \left|\sum_{m,n,o,p=1,...,d_\delta}\frac{\partial^4l(\theta^+;\obs_{1:\n}) }{\partial \delta_m\partial\delta_n\partial \delta_o\partial\delta_p} - \frac{\partial^4l(\theta_0;\obs_{1:\n}) }{\partial \delta_m\partial\delta_n\partial \delta_o\partial\delta_p}\right| \\
     &+  \|\lambda\|^4  \left|\sum_{m,n,o,p=1,...,d_\lambda}\frac{\partial^4l(\theta^+;\obs_{1:\n}) }{\partial \lambda_m\partial\lambda_n\partial \lambda_o\partial\lambda_p} - \frac{\partial^4l(\theta_0;\obs_{1:\n}) }{\partial \lambda_m\partial\lambda_n\partial \lambda_o\partial\lambda_p}\right| 
\end{align*}

and then: 

\begin{align*}
    |R_\n(\theta)|\leq \mathcal{O}_p(1)\|t_\n(\theta)\|^2 (o_p(1) &+ \frac{1}{\n}\left|\sum_{m,n,o,p=1,...,d_\lambda}\frac{\partial^4l(\theta^+;\obs_{1:\n}) }{\partial \lambda_m\partial\lambda_n\partial \lambda_o\partial\lambda_p} - \frac{\partial^4l(\theta_0;\obs_{1:\n}) }{\partial \lambda_m\partial\lambda_n\partial \lambda_o\partial\lambda_p}\right|  \\
    & +  \frac{1}{\n}\left|\sum_{m,n,o,p=1,...,d_\delta}\frac{\partial^4l(\theta^+;\obs_{1:\n}) }{\partial \delta_m\partial\delta_n\partial \delta_o\partial\delta_p} - \frac{\partial^4l(\theta_0;\obs_{1:\n}) }{\partial \delta_m\partial\delta_n\partial \delta_o\partial\delta_p}\right| )
    \end{align*} 
    
where $\theta^+= \theta_0 + t(\theta-\theta_0)$ for some $0<t<1$.

To show that the last two terms tend to zero we proceed as follows : 

\begin{align*}
    \frac{1}{\n}|\sum_{m,n,o,p=1,...,d_\lambda}\frac{\partial^4l(\theta^+;\obs_{1:\n}) }{\partial \lambda_m\partial\lambda_n\partial \lambda_o\partial\lambda_p}& - \frac{\partial^4l(\theta_0;\obs_{1:\n}) }{\partial \lambda_m\partial\lambda_n\partial \lambda_o\partial\lambda_p}|=\\
    &\frac{1}{\n}|\sum_{m,n,o,p=1,...,d_\lambda}\frac{\partial^4l(\theta^+;\obs_{1:\n}) }{\partial \lambda_m\partial\lambda_n\partial \lambda_o\partial\lambda_p} - \E\left[\sum_{m,n,o,p=1,...,d_\lambda}\frac{\partial^4l(\theta^+;\obs_1) }{\partial \lambda_m\partial\lambda_n\partial \lambda_o\partial\lambda_p}\right]\\
    & + \E\left[\sum_{m,n,o,p=1,...,d_\lambda}\frac{\partial^4l(\theta^+;\obs_{1:\n}) }{\partial \lambda_m\partial\lambda_1\partial \lambda_o\partial\lambda_p}\right]- \E\left[\frac{\partial^4l(\theta_0;\obs_1) }{\partial \lambda_m\partial\lambda_n\partial \lambda_o\partial\lambda_p}\right] \\
    &+ \E\left[ \frac{\partial^4l(\theta_0;\obs_1) }{\partial \lambda_m\partial\lambda_n\partial \lambda_o\partial\lambda_p}\right] - \frac{\partial^4l(\theta_0;\obs_{1:\n}) }{\partial \lambda_m\partial\lambda_n\partial \lambda_o\partial\lambda_p}|
\end{align*}

We then apply triangular inequality to separate the 3 terms. The first and third terms are empirical means of centered random variables with bounded variances (assumption \eqref{ass:regularity 0-4}$v)$). Therefore we can use each time Chebychev's inequality to obtain a weak law of large number and obtain the consistency toward $0$. For the second term, 

\[|\sum_{m,n,o,p=1,...,d_\lambda}\frac{\partial^4l(\theta^+;\obs_1) }{\partial \lambda_m\partial\lambda_1\partial \lambda_o\partial\lambda_p} -\sum_{m,n,o,p=1,...,d_\lambda}\frac{\partial^4l(\theta;\obs_1) }{\partial \lambda_m\partial\lambda_1\partial \lambda_o\partial\lambda_p} |\leq 2C\underset{\theta\in\Theta}{\sup}\|\nabla^4_\theta l(\theta;\obs_1)\|\] 

where $C$ is the nonnegative constant that appears in the equivalence between the L1 and L2 norm. When we evaluate at $\theta = \emv\theta$,  $\theta^+ = \theta_0 + t(\emv\theta-\theta_0)\rightarrow\theta_0$ in probability, as $\n\rightarrow +\infty$. And by continuity of $\nabla^4_\theta l(\cdot;\obs_1)$, and dominated convergence, the second term also converges toward $0$. Finally we obtain :

\[|R_\n(\emv{\theta})| \leq o_p(1)\|t_\n(\emv{\theta})\|^2\]

And then:
    $0\leq l(\emv{\theta};\obs_{1:\n})-l(\theta_0;\obs_{1:\n})$\\
    $\leq \|\tilde{S}_\n(\theta_0)\|\|t_\n(\emv{\theta})\|-\frac{1}{2}\|t_\n(\emv{\theta})\|^2_{\tilde{I}_\n(\theta_0)} + o_p(\|t_\n(\emv{\theta})\|^2)$\\
    $\leq  \|\tilde{S}_\n(\theta_0)\|\|t_\n(\emv{\theta})\|-\frac{1}{2}(o_p(1) + a)\|t_\n(\emv{\theta})\|^2$\\
where 

\[a = \underset{\n>n_0}{\inf}\underset{x\neq0}{\inf}\frac{\|x\|^2_{\tilde{I}_\n(\theta_0)}}{\|x\|^2}\]

where $\|x\|_A$ stands for $x^TAx$, with $A$ being a positive definite symmetric matrix. 

By taking $n_0$ large enough so that for every $\n>n_0$, $\tilde{I}_\n(\theta_0)\succ 0$ (assumption \eqref{ass:Itilde}) we have that $0<a<+\infty$. The last inequality, shows that for $\n$ large enough, this polynomial of degree 2 in $\|t_\n(\emv{\theta})\|$ is upper bounded (dominant coefficient negative) and lower bounded by 0. Which shows that

\begin{equation*}
    t_\n(\emv{\theta}) = \mathcal{O}_p(1)
\end{equation*}

which concludes the proof, and :

\begin{equation}\label{eq:reste}
    R_\n(\emv{\theta})=o_p(1)\mathcal{O}_p(1)=o_p(1)
\end{equation}

which is fundamental for the proof of theorem \eqref{prop:LRTbootsasym}

\subsection{Proof of theorem \ref{prop:LRTbootsasym}}

Now that we derived the expression of $LRT_\infty$, it remains to show that  the bootstrap statistic also converges weakly in probability to this random variable. To do so we first derive a bootstrap quadratic approximation as in \eqref{eq:quadratic2}, where the expansion is done around $\emvB\theta$, as it is the true parameter of the bootstrap data. We obtain that : 

\begin{equation}\label{eq:quadratic2}
\begin{split}
    l(\theta;y^*_{1:\n}) - l(\emvB{\theta};\obs_{1:\n}^*) = \frac{1}{2}&Z_\n^*(\emvB\theta)^T \tilde{I}^*_\n(\emvB\theta)Z_\n^*(\emvB\theta)\\ 
    &-  \frac{1}{2}\left(t_\n^*(\theta) - Z_\n^*(\emvB\theta)\right)^T\tilde{I}^*_\n(\emvB\theta)\left(t_\n^*(\theta) - Z_\n^*(\emvB\theta)\right)+ R_\n^*(\theta)
    \end{split}
\end{equation}
where the exponent $^*$ stands for "evaluated on the bootstrap data". 
The following proof will follow three main steps. 

First we show that the bootstrap version of $\tilde{I}_\n(\theta_0)$ and of $\tilde{S}_\n(\theta_0)$ have the correct conditional limiting distribution which means that : 

\begin{table}[h]
    \centering
    \begin{equation}\label{eq:objectif}
    \left\{\begin{tabular}{ll}
          $\tilde{I}_\n^*(\emvB{\theta})-\tilde{I}_\n(\theta_0) = o_{p^*}(1)$ \\
          $\forall t\in \mathds{R}^{d_{\tilde{I}}}\quad \pr^*\{\tilde{S}^*_\n(\emvB\theta)<t\}-\pr\{\tilde{S}_\n(\theta_0)<t\} = o_{p}(1)$
    \end{tabular}\right.
    \end{equation}
    \label{tab:my_label}
\end{table}

Second, we show that the new terms in the quadratic approximation that are supposed to be null converge toward $0$ thanks to the shrinkage parameter. 

Finally we show that the rest in the bootstrap quadratic approximation is a $o_{p^*}(1)$.

We first consider the case with no nuisance parameters i.e. $\theta = (\psi, \lambda) $ to lighten the notations and work in two steps.\vspace{0.5 cm}

We first consider $\tilde{I}_\n(\emvB{\theta})$. We write :
  
  \[\left[\tilde{I}_\n(\emvB{\theta})\right]_{m,n} = \frac{1}{\n}\sum_{i=1}^{\n}h_{m,n}^{(i)*}(\emvB{\theta})\]
  
where $h_{m,n}^{(i)}(\emvB{\theta})$ is of the form:

\[h_{m,n}^{(i)*}(\emvB{\theta}) = c_{m,n}\frac{\partial^k\log f(\obs_i;\emvB{\theta})}{\partial \theta_{i_1}...\partial\theta_{i_k}}\] 

with $c_{m,n}\in\mathbb{R}$, $k\in\{2,4\}$, $1\leq i_1,...,i_k\leq q_\lambda +q_\psi$.

We want to show that:

\[\frac{1}{\n}\sum_{i=1}^{\n}h_{m,n}^{(i)*}(\emvB{\theta}) - \E\{h_{m,n}^{(1)}(\theta_0)\}=o_{p^*}(1)\]

To show that we decompose the difference:

\[\frac{1}{\n}\sum_{i=1}^{\n}h_{m,n}^{(i)*}(\emvB{\theta})-\E\left[h_{m,n}^{(1)}(\theta_0)\right]\]

as:

\[\overbrace{\frac{1}{\n}\sum_{i=1}^{\n}h_{m,n}^{(i)*}(\emvB{\theta})-\E^*\left[h_{m,n}^{(1)*}(\emvB{\theta})\right]}^{(T1)} + \overbrace{\E^*[h_{m,n}^{(1)*}(\emvB{\theta})] - \E^*[h_{m,n}^{(1)*}(\theta_0)]}^{(T2)} + \overbrace{\E^*[h_{m,n}^{(1)*}(\theta_0)] -\E\left[h_{m,n}^{(1)}(\theta_0)\right]}^{(T3)}\]

The term $(T1)$ is a sum of centered random variables, we can apply Chebychev's inequality to have the convergence towards $0$ (using assumption \eqref{ass:regularity 0-4} to have that the variance is $\mathcal{O}_p(1)$.\\\\
The term $(T2)$ is controlled as follows :

\begin{align*}
    |\E^*[(h_{m,n}^{(1)*}(\emvB{\theta})-h_{m,n}^{(1)*}(\theta_0))]|&\leq \E^*\left[|(h_{m,n}^{(1)*}(\emvB{\theta})-h_{m,n}^{(1)*}(\theta_0))|\right]\\
    & \leq \underset{\theta\in\Theta}{\sup}\E_\theta\left[ |(h_{m,n}^{(1)*}(\emvB{\theta})-h_{m,n}^{(1)*}(\theta_0))|\right] \\
\end{align*}

as $|(h_{m,n}^{(1)*}(\emvB{\theta})-h_{m,n}^{(1)*}(\theta_0))|\leq2\underset{\theta\in\Theta}{\sup} |h_{m,n}^{(1)}(\theta)|$ almost surely, using  assumption \ref{ass:regularity 0-4} and thanks to the consistency of $\emvB{\theta}$ we can use the dominated convergence theorem  to prove the convergence in probability toward $0$.\\\\
Finally we deal with $(T3)$ as follows :
  
\begin{align*}
    |\E^*[h_{m,n}^{(1)*}(\theta_0)] -\E\left[h_{m,n}^{(1)}(\theta_0)\right]| &= |\int h_{m,n}^{(1)*}(\theta_0) \{f(\obs;\emvB\theta)-f(\obs;\theta_0)\}dy|\\
    &\leq \int |h_{m,n}^{(1)*}(\theta_0)| |f(\obs;\emvB\theta)-f(\obs;\theta_0)|dy
\end{align*}

To show that this term tends toward $0$ we use another time the equation \eqref{eq:outil1}, first we use a taylor expansion, there exist $\theta^+$ between $\theta_0$ and $\emvB\theta$ such that : 

\begin{align*}
    f(\obs;\emvB\theta)-f(\obs;\theta_0) &= (\emvB\theta-\theta_0)^T\nabla_\theta f(\obs;\theta^+)\\&= (\emvB\theta-\theta_0)^T\nabla_\theta \log f(\obs;\theta^+) f(\obs;\theta^+) 
\end{align*}

therefore, 

\begin{align*}
    |f(\obs;\emvB\theta)-f(\obs;\theta_0)|&\leq \|\emvB\theta-\theta_0\|\|\nabla_\theta \log f(\obs;\theta^+)\| f(\obs;\theta^+) 
\end{align*}

and, 

\begin{align*}
     |\E^*[h_{m,n}^{(1)*}(\theta_0)] -&\E\left[h_{m,n}^{(1)}(\theta_0)\right]|\leq  \|\emvB\theta-\theta_0\|\int |h_{m,n}^{(1)*}(\theta_0)|\|\nabla_\theta \log f(\obs;\theta^+)\| f(\obs;\theta^+) dy\\
     &\leq \|\emvB\theta-\theta_0\| \int \frac{1}{2} \left\{ |h_{m,n}^{(1)*}(\theta_0)|^2 +\|\nabla_\theta \log f(\obs;\theta^+)\|^2\right\}f(\obs;\theta^+) dy
\end{align*}

Thanks to assumption \eqref{ass:regularity 0-4}, as $\emvB\theta-\theta_0=o_p(1)$ this last term is $o_p(1)$ as $\n\rightarrow+\infty$.\vspace{0.5cm}

  We then consider $\tilde{S}_\n^*(\emvB{\theta})$, which is $\sqrt{\n}$ times a sum of (conditionally) independent, centered, with finite variance  random variables. Therefore, by proving that :
  
\begin{equation}\label{eq:varboot iid}
    \E^*[\tilde{S}_\n^*(\emvB{\theta})\tilde{S}_\n^*(\emvB{\theta})^T]-\E\left[\tilde{S}_1(\theta_0)^T\tilde{S}_1(\theta_0)^T\right] = o_{p}(1)
\end{equation}

and applying a multivariate version of the conditional central limit theorem of \cite{bulinski2017conditional}, it will conclude the second part of (13). Our third moment condition, and the consistency of the variance matrix of the score is much stronger then the Lindberg Feller conditions.

For each $m,n$, $\left[\tilde{S}_\n^*(\emvB{\theta})\tilde{S}_\n^*(\emvB{\theta})^T\right]_{m,n}$ can also be written as $\frac{1}{\n}\sum_{i=1}^{\n}h_{m,n}^{(i)*}(\emvB{\theta})$, where $h_{m,n}^{(i)*}(\emvB{\theta}) = [\tilde{S}_\n^*(\emvB{\theta})]_m\times [\tilde{S}_\n^*(\emvB{\theta})]_n $ . Due to assumption \eqref{ass:regularity 0-4}$(ii)$--$(iii)$, and the elementary equation \eqref{eq:outil1}, $\E\left[h_{m,n}^{(i)*}(\emvB{\theta})^{3/2}\right]<+\infty$.   

In order to prove \eqref{eq:varboot iid}, we once again split the sum : 

\begin{align*}
    \E^*[h_{m,n}^{(1)*}(\emvB\theta)]-&\E\left[h_{m,n}^{(1)}(\theta_0)\right] =  \\
    \overbrace{\E^*[h_{m,n}^{(1)*}(\emvB\theta)] -\E^*\left[h_{m,n}^{(1)*}(\theta_0)\right]}^{(S1)} &+
    \overbrace{\E^*[h_{m,n}^{(1)*}(\theta_0)] -\E\left[h_{m,n}^{(1)}(\theta_0)\right]}^{(S2)}
\end{align*}

We first deal with $(S1)$ :
\begin{align*}
    |\E^*[h_{m,n}^{(1)*}(\emvB\theta)] -\E^*\left[h_{m,n}^{(1)*}(\theta_0)\right]| &\leq \underset{\theta\in\Theta}{\sup} \quad|\E_\theta[h_{m,n}^{(1)*}(\emvB\theta)-h_{m,n}^{(1)*}(\theta_0)]|\\
    &\leq \underset{\theta\in\Theta}{\sup}\quad \E_\theta[|h_{m,n}^{(1)*}(\emvB\theta)-h_{m,n}^{(1)*}(\theta_0)|]\\&\leq 
    2\underset{\theta\in\Theta}{\sup} \quad\E_\theta[\underset{\theta'\in\Theta}{\sup}|h_{m,n}^{(1)*}(\theta')|] \\
    &<+\infty
    \end{align*}
    
Which enables, to apply dominated convergence to the first term, as $\emv\theta$ is consistent.

For $(S2)$ we proceed as follows :

\begin{align*}
    |\E^*[h_{m,n}^{(1)*}(\theta_0)] -\E\left[h_{m,n}^{(1)}(\theta_0)\right]| &= |\int h_{m,n}^{(1)*}(\theta_0) \{f(\obs;\emvB\theta)-f(\obs;\theta_0)\}dy|\\
    &\leq \int |h_{m,n}^{(1)*}(\theta_0)| |f(\obs;\emvB\theta)-f(\obs;\theta_0)|dy
\end{align*}

To show that this term tends toward $0$ we use the same reasonning as before, first we use a Taylor expansion: there exist $\theta^+$ between $\theta_0$ and $\emvB\theta$ such that : 

\begin{align*}
    f(\obs;\emvB\theta)-f(\obs;\theta_0) &= (\emvB\theta-\theta_0)^T\nabla_\theta f(\obs;\theta^+)\\&= (\emvB\theta-\theta_0)^T\nabla_\theta \log f(\obs;\theta^+) f(\obs;\theta^+) 
\end{align*}

therefore, 

\begin{align*}
    |f(\obs;\emvB\theta)-f(\obs;\theta_0)|&\leq \|\emvB\theta-\theta_0\|\|\nabla_\theta \log f(\obs;\theta^+)\| f(\obs;\theta^+) 
\end{align*}

and, 

\begin{align*}
     |\E^*[h_{m,n}^{(1)*}(\theta_0)] -&\E\left[h_{m,n}^{(1)}(\theta_0)\right]|\leq  \|\emvB\theta-\theta_0\|\int |h_{m,n}^{(1)*}(\theta_0)|\|\nabla_\theta \log f(\obs;\theta^+)\| f(\obs;\theta^+) dy\\
\end{align*}

We can't directly use equation \eqref{eq:outil1} here because $h_{m,n}^{(1)}(\theta) $ doesn't admit second order moments ($\tilde{S}_\n(\theta)$ admits third order moments). Thanks to Holder's inequality using $p=\frac{3}{2}$ and $q=3$ we have that : 

\begin{align*}
    &\|\emvB\theta-\theta_0\|\int |h_{m,n}^{(1)*}(\theta_0)|\|\nabla_\theta \log f(\obs;\theta^+)\| f(\obs;\theta^+) dy \\ 
    &\leq \|\emvB\theta-\theta_0\| \left(\int\|\nabla_\theta \log f(\obs;\theta^+)\|^3f(\obs;\theta^+) dy\right)^{\frac{1}{3}}\left(\int |h_{m,n}^{(1)*}(\theta_0)|^{\frac{3}{2}}f(\obs;\theta^+) dy\right)^{\frac{2}{3}}\\& = o_p(1)
\end{align*}

the last inequality holds thanks to assumption \eqref{ass:regularity 0-4}$(ii)$--$(iii)$ that enables to state that the two integrals are finite. That concludes the proof that \eqref{eq:varboot iid} holds.\vspace{0.5cm}

  We now deal with the nuisance parameters. The proof starts the same way as  before. The issue is that the odd derivatives with respect to the parameter $\delta$ are not $0$ when evaluated at $\emvB{\theta}$ because $\emvB{\delta}\neq 0$. We recall that $\theta=(\psi, \delta,\lambda)$ and $\emvB{\theta}=(\emvB{\psi},\emvB{\delta}, 0)$
  
To lighten the calculations we write $dx^* = x-\emvB{x}$,  for any quantity $x$. 

After expanding the likelihood  as \eqref{eq:quadratic2}, new terms appear in $R_\n^*(\theta)$: the odds order derivatives with respect to $\delta$ which would be $0$ if $\emvB{\delta}=0$. 

We want to show that :
 $d\delta^{*T}\nabla_\delta l(\emvB{\theta};\obs_{1:\n}^*)=o_{p^*}(1)$ and  $\sum_{ijk}d\delta_i^* d\delta_j^* d\delta_k^*\frac{\partial^3l(\emvB{\theta};\obs_{1:\n}^*)}{\partial\delta_i\partial\delta_j\partial\delta_k}= o_{p^*}(1)$.

As said before the issue comes from the fact that the boostrap parameter of $\delta$ is not $0$ . Indeed the "good" bootstrap parameter would be $\emvB{u}=(\emvB{\psi},0,0) = \emvB{\theta}-(0,\emv\delta,0)$ ( the bootstrap parameter that we would use if we knew where were located the nuisance parameters) .

We are now going to expand the terms around $\emvB{u}$, so that the odds order derivatives evaluated at $\emvB{u}$ will be zero. And we will use the fact that $\emvB\delta$ converges very fast to zero.

We write $\theta^+ = t\emvB{\theta} + (1-t)\emvB{u}$: 

\begin{align*}
    d\delta^{*T}\nabla_\delta l(\emvB{\theta};\obs_{1:\n}^*) & = d\delta^{*T}\left(0 + \nabla_{\delta}^2 l_\n^*(\emvB{u})\emvB{\delta} + \sum_{ij}\emvB{\delta}_i\emvB{\delta}_j\times 0 + \sum_{ijk}\emvB{\delta}_i\emvB{\delta}_j\emvB{\delta}_k\frac{\partial^4 l(\theta^+;\obs_{1:\n}^*)}{\partial \delta_i\partial \delta_j\partial \delta_k\partial \delta}\right)
\end{align*}
the first non zero term is a centered random variable with finite variance (assumption \eqref{ass:regularity 0-4}$(iii)$), therefore by the central limit theorem it is $\mathcal{O}_{p^*}(\sqrt{\n})$, and the last term is $\mathcal{O}_{p^*}(1)$ by the law of large number. Therefore: 
\begin{align*}
    |d\delta^{*T}\nabla_\delta l(\emvB{\theta};\obs_{1:\n}^*)|&\leq \|d\delta^*\|\left(\mathcal{O}_p(\sqrt{\n})o_p(\n^{-\frac{1}{4}}) + \mathcal{O}_p(\n)o_p(\n^{-\frac{3}{4}})\right)\\
    &\leq \|d\delta^*\|o_p(\n^{\frac{1}{4}})
\end{align*}
  The same way we have:
\begin{align*}
    |\sum_{ijk}d\delta_i^* d\delta_j^* d\delta_k^*\frac{\partial^3l(\emvB{\theta};\obs_{1:\n}^*)}{\partial\delta_i\partial\delta_j\partial\delta_k}|& = 0+|\emvB{\delta}^T  \sum_{ijk}d\delta_i^* d\delta_j^* d\delta_k^*\frac{\partial^4l(\theta^+;\obs_{1:\n}^*)}{\partial\delta\partial\delta_i\partial\delta_j\partial\delta_k}| \\
    &\leq o_{p^*}(\n^{\frac{-1}{4}})\|d\delta^*\|^3 \times \mathcal{O}_{p^*}(\n)\\
    &\leq \|d\delta^*\|^3 o_{p^*}(\n^{\frac{3}{4}})
\end{align*}
By using the same reasoning as in section \eqref{sec:Alrtasym} for the derivation of the likelihood ratio statistic, evaluating the expansion of the bootstrap likelihood at the bootstrap maximum likelihood estimator (restricted or not) $\hat{\theta}^*$, we have that almost surely 
\begin{align*}
    0&\leq l(\hat{\theta}^*;\obs_{1:\n}^*) - l(\emvB{\theta};\obs_{1:\n}^*)
    \\&\leq \|\tilde{S}_\n^*(\emvB{\theta})\|\|t_\n(\hat{\theta}^*)\|-\frac{1}{2}\|t_\n(\hat{\theta}^*)\|^2_{\tilde{I}^*_\n(\emvB{\theta})} + R^*_\n(\hat{\theta}^*)\\
&\leq\|\tilde{S}_\n^*(\emvB{\theta})\|\|t_\n(\hat{\theta}^*)\|-\frac{1}{2}(o_{p^*}(1) + a^*)\|t_\n(\hat{\theta}^*)\|^2 + o_{p^*}(1)(\|t_\n(\hat{\theta}^*)\|^{\frac{1}{2}}+ \|t_\n(\hat{\theta}^*)\|^{\frac{3}{2}})\\
\end{align*}
Even if this quantity is no longer a polynomial, the dominant term remain the same, therefore this quantity is lower bounded by $0$ and upper bounded in probability as a upper bounded function of $\|t_\n(\hat{\theta}^*)\|$. Which implies that $\|t_\n(\hat{\theta}^*)\|=\mathcal{O}_{p^*}(1)$ (in this proof we don't show that it is not a $o_{p^*}(1)$ but it is not important here as we already showed that the bootstrap score and the bootstrap FIM converge toward the correct limit). But the important is that we showed that $R_\n^*(\hat{\theta}^*)=o_{p^*}(1)$ which conclude the proof. \\\\
 
\subsection{Proof of proposition \ref{lem:shrink}}
We recall that $\emv\theta = (\emv\psi,\emv\delta,\emv\lambda)=  \arg \underset{\theta\in\Theta}{\max}\quad \lnz$, and $(c_\n)$ is a sequence defined as in proposition \eqref{lem:shrink} . 

Consider $\emvB{\theta} = (\emvB{\psi}, \emvB{\delta},\emvB{\lambda})$ such that $\forall k=1,..,d_\psi$ $\psi^*_{\n,k} = \hat{\psi}_{\n,k}\ \mathds{1}(\hat{\psi}_{\n,k}>c_\n)$, $\forall k=1,..,d_\delta$ $\delta^*_{\n,k} = \hat{\delta}_{\n,k}\ \mathds{1}(\hat{\delta}_{\n,k}>c_\n)$ and  $\emvB{\lambda} = 0_{d_\lambda}$. 

The proof of this proposition follows exactly the lines of \cite{cavaliere2020bootstrap} lemma 1. First the fact that $\n^{1/4}c_\n\rightarrow +\infty$ as $\n\rightarrow +\infty$ implies also that $\sqrt{\n}c_\n\rightarrow +\infty$. \vspace{0.5cm}

Let us establish a technical result that will then be applied to our proposition. Let $(x_\n)$ a real valued random sequence and $x_0\in\mathds{R}$ such that $r_\n(x_\n-x_0) = O_p(1)$, for $r_\n$ being whether $\sqrt{\n}$ of $N^{1/4}$.

If $x_0 = 0$, 
\begin{equation*}
    \pr(x_\n>c_\n) = \pr(r_\n x_\n>r_\n c_\n) = \pr\{O_p(1)>r_\n c_\n\} = o(1)
\end{equation*}
The last equality holds because $r_\n c_\n\rightarrow +\infty$. Therefore $\mathds1 (x_\n>c_\n) = o_p(1)$ and finally $r_\n x_\n \mathds1 (x_\n>c_\n) = O_p(1)o_p(1)=o_p(1)$. 

If $x_0\neq 0$, 
\[\pr(|x_\n|>c_\n)  = \pr(|x_\n -x_0+x_0|>c_\n) \geq  \pr\{||x_\n-x_0|-|x_0||>c_\n\} \]

\begin{equation*}
    \pr\{||x_\n-x_0|-|x_0||>c_\n\} = \pr\{|x_0|-|x_\n-x_0|>c_\n\} + \pr\{|x_\n-x_0| - |x_0|>c_\n\}
\end{equation*}

First, $|x_\n-x_0|+c_\n=o_p(1)$ and $|x_0|>0$ therefore $\pr\{|x_0|-|x_\n-x_0|>c_\n\}\rightarrow 1$ as $\n\rightarrow +\infty$.  And $r_\n(|x_0|+c_\n)\rightarrow+\infty$ so $\pr\{|x_\n-x_0| - |x_0|>c_\n\}\rightarrow 0$ as $\n\rightarrow+\infty$.  Therefore : 
\begin{equation}\label{eq:indic}
    \mathds1 (x_\n>c_\n)-1 = o_p(1)
\end{equation}

Finally, 

\begin{align*}
    r_\n\{x_\n\mathds1(x_\n>c_\n) - x_0\} &=  r_\n(x_\n - x_0)\mathds1(x_\n>c_\n) - x_0\mathds1(x_\n\leq c_\n)\\ 
    &=  r_\n(x_\n - x_0)\mathds1(x_\n>c_\n) - x_0\{1-\mathds1(x_\n> c_\n)\}\\&= O_p(1) +o_p(1)
\end{align*}
using equation \eqref{eq:indic}, it concludes with Slutsky's theorem that $r_\n\{x_\n\mathds1(x_\n>c_\n) - x_0\}$ and $r_\n(x_n-x_0)$ have the same limiting distribution.

Applying this result to $x_\n=\emv\delta$ and $r_\n=\n^{1/4}$ we have that $\emvB\delta = o_p(1)$. the same holds for $\emv\psi$ : $\sqrt{\n}(\emv\psi-\psi_0)=O_p(1)$ (if $\psi_0=0$ then $\sqrt{\n}\emv\psi=o_p(1)$ ).  

Finally  it is obvious that $\emvB\theta\in\Theta_0$ as $\lambda =0_{d_\lambda}$. Which concludes the proof.

\subsection{Proof of proposition \ref{prop:mle not iid}}

We verify that our hypothesis imply the conditions required in \cite{hoadley1971asymptotic}. \\
We show easily that the assumptions C(1), C(2), C(3'),C(4'), C(5) are verified.
Assumptions C(1)--(2) are verified with assumption \eqref{ass:regularity1}. Assumption C(3') is weaker than assumption \eqref{ass:notiid}. C(4') is equivalent to assumption \eqref{ass:not iid unicity}. C(5) is verified as we the continuity of the likelihood with respect to $\theta$, for every $\obs$ and the measurability with respect to $\obs$ for every $\theta$. The result is then discussed for instance in \cite*{gine2021mathematical} exercise 7.2.3. 

\subsection{Proof of theorem \ref{prop:LRTbootsasym not iid}}

This proof is very similar to the one of theorem \eqref{prop:LRTbootsasym}.

First we have to derive the asymptotic distribution of the likelihood ratio test statistic. We start from the quadratic expansion \eqref{eq:quadratic2}. We first show that $\tilde{I}_\n(\theta_0)$ converges in probability toward a non random matrix $\tilde{I}(\theta_0)$.

As in the proof of theorem \eqref{prop:LRTbootsasym} we write
\[\left[\tilde{I}_\n(\theta_0)\right]_{m,n} = \frac{1}{\n}\sum_{i=1}^{\n}h_{m,n}^{(i)}(\theta_0)\]

where $h_{m,n}^{(i)}(\theta_0)$ is of the form:

\[h_{m,n}^{(i)}(\theta_0) = c_{m,n}\frac{\partial^k\log f_i(\obs_i;\theta_0)}{\partial \theta_{i_1}...\partial\theta_{i_k}}\] 

with $c_{m,n}\in\mathbb{R}$, $k\in\{2,4\}$, $1\leq i_1,...,i_k\leq d_\psi+d_\lambda+d_\delta$.

As a consequence of assumption \eqref{ass:notiid}, using Chebychev's inequality : 

\[\frac{1}{\n}\sum_{i=1}^{\n}|h_{m,n}^{(i)}(\theta_0)-\E[h_{m,n}^{(i)}(\theta_0)]| = o_p(1)\]

which enables to define $\tilde{I}(\theta_0)  = \left[\underset{\n\rightarrow +\infty}{\lim}\frac{1}{\n}\sum_{i=1}^\n\E[h_{m,n}^{(i)}(\theta_0)]\right]_{m,n}$ which is a nonrandom matrix that is supposed to be positive definite (assumption \eqref{ass:Itilde}). 
Furthermore, 
\begin{align*}
    \frac{1}{\n}\sum_{i=1}^\n\E\left[|h_{m,n}^{(i)}(\theta_0)|\right]&\leq  \frac{1}{\n}\sum_{i=1}^\n\underset{i\in\mathds{N}}{\sup}\quad\E\left[|h_{m,n}^{(i)}(\theta_0)|\right]\\ 
    &\leq \underset{i\in\mathds{N}}{\sup}\quad\E\left[|h_{m,n}^{(i)}(\theta_0)|\right] \\&< +\infty 
\end{align*}
which holds for every $\n\geq 0$. This last inequality enables to invert the sum and the integral : 

\begin{align*}
    \left[\tilde{I}(\theta_0)\right]_{m,n} &=  \underset{\n\rightarrow +\infty}{\lim}\frac{1}{\n}\sum_{i=1}^\n\E[h_{m,n}^{(i)}(\theta_0)] \\
    &= \E\left[\underset{\n\rightarrow +\infty}{\lim}\frac{1}{\n}\sum_{i=1}^\n h_{m,n}^{(i)}(\theta_0)\right] \\
    &= \underset{\n\rightarrow +\infty}{\lim}\frac{1}{\n}\sum_{i=1}^\n h_{m,n}^{(i)}(\theta_0)
\end{align*}

where the last equality holds as we consider a non random quantity.

We consider now $\tilde{S}_\n(\theta_0)$ which is a sum of centered random variables with finite variances, we want to apply theorem 6.5 of \cite{hansen2022econometrics}. Assumption \eqref{ass:regularity 0-4}$(ii)$--$(iii)$ and assumption \eqref{ass:notiid} enables to state that :
\[ \underset{\n\rightarrow +\infty}{\lim}\E\left[ \tilde{S}_\n(\theta_0)\tilde{S}_\n(\theta_0)^T \right] < +\infty\]

which is a direct consequence of theorem A.5 of \cite{hoadley1971asymptotic}.Furthermore,  still thanks to  assumption \eqref{ass:regularity 0-4}$(ii)$--$(iii)$ and assumption \eqref{ass:notiid} equation $(6.3)$ in \cite{hansen2022econometrics} theorem 6.5 is verified for $\delta = 1 $, and therefore $\tilde{S}_\n(\theta_0)$ is $O_p(1)$ and converges in distribution toward a random variable that we call $U(\theta_0)$. 

The next step of the proof is to prove \eqref{eq:objectif}. 

We first deal with $\tilde{I}_\n^*(\emvB\theta)$, we still write : 

 \[\left[\tilde{I}_\n(\emvB{\theta})\right]_{m,n} = \frac{1}{\n}\sum_{i=1}^{\n}h_{m,n}^{(i)*}(\emvB{\theta})\]

We proceed as in the proof of theorem \eqref{prop:LRTbootsasym}, and we split :

\begin{align*}
    \frac{1}{\n}\sum_{i=1}^{\n}h_{m,n}^{(i)*}(\emvB{\theta})-\left[\tilde{I}(\theta_0)\right]_{m,n} &= \frac{1}{\n}\sum_{i=1}^{\n}h_{m,n}^{(i)*}(\emvB{\theta}) - \underset{\n\rightarrow +\infty}{\lim}\frac{1}{\n}\sum_{i=1}^\n\E[h_{m,n}^{(i)}(\theta_0)] \\
    &=  \frac{1}{\n}\sum_{i=1}^{\n}h_{m,n}^{(i)*}(\emvB{\theta}) -\frac{1}{\n}\sum_{i=1}^\n\E[h_{m,n}^{(i)}(\theta_0)] + o(1)
\end{align*}

as 

\begin{equation*}
\begin{split}
\overbrace{\frac{1}{\n}\sum_{i=1}^{\n}h_{m,n}^{(i)*}(\emvB{\theta})-\E^*\left[h_{m,n}^{(i)*}(\emvB{\theta})\right]}^{(U1)} + &\overbrace{\frac{1}{\n}\sum_{i=1}^{\n}\E^*[h_{m,n}^{(i)*}(\emvB{\theta})] - \E^*[h_{m,n}^{(i)*}(\theta_0)]}^{(U2)} \\
& + \overbrace{\frac{1}{\n}\sum_{i=1}^{\n}\E^*[h_{m,n}^{(i)*}(\theta_0)] -\E\left[h_{m,n}^{(i)}(\theta_0)\right]}^{(U3)} + o(1)
\end{split}
\end{equation*}

The term $(U1)$ is a sum of centered random variables with finite variance uniformly bounded over $i \in \mathds{n}$ and therefore is $o_{p^*}(1)$. 

We deal with $(U2)$ as before : 

\begin{align*}
    |\frac{1}{\n}\sum_{i=1}^{\n}\E^*[h_{m,n}^{(i)*}(\emvB{\theta})] - \E^*[h_{m,n}^{(i)*}(\theta_0)]| &\leq \frac{1}{\n}\sum_{i=1}^{\n}\E^*[|h_{m,n}^{(i)*}(\emvB{\theta}) - h_{m,n}^{(i)*}(\theta_0) |] \\
    &\leq \underset{i\in \mathds{N}}{\sup} \quad \underset{\theta\in\Theta}{\sup} \quad \E_{\theta}\left[ |h_{m,n}^{(i)*}(\emvB{\theta}) - h_{m,n}^{(i)*}(\theta_0) |\right]
\end{align*}

and for every $i$,  $|h_{m,n}^{(i)*}(\emvB{\theta}) - h_{m,n}^{(i)*}(\theta_0) |\leq 2 \underset{\theta'\in\Theta}{\sup}|h_{m,n}^{(i)*}(\theta')|$, thanks to assumption \eqref{ass:notiid}, we can apply dominated convergence. 

For the term $(U3)$, we apply the exact same reasoning as in the proof of theorem \eqref{prop:LRTbootsasym} to show that 

\begin{align*}
    |\frac{1}{\n}\sum_{i=1}^{\n}\E^*[h_{m,n}^{(i)*}(\theta_0)] -\E\left[h_{m,n}^{(i)}(\theta_0)\right]|
\end{align*}

is almost surely smaller than 

\[\|\emvB\theta-\theta_0\|\quad \underset{i\in\mathds{N}}{\sup}\quad\int \frac{1}{2} \left\{ |h_{m,n}^{(i)*}(\theta_0)|^2 +\|\nabla_\theta \log f_i(\obs;\theta^+)\|^2\right\}f_i(\obs;\theta^+) dy\]

which is almost surely smaller than 

\[ \frac{\|\emvB\theta-\theta_0\|}{2} \left\{ \underset{\theta^+\in\Theta}{\sup} \E_{\theta^+}\left[\underset{\theta\in\Theta}{\sup}\quad |h_{m,n}^{(i)}(\theta)|^2\right]  + \underset{\theta^+\in\Theta}{\sup} \E_{\theta^+}\left[\underset{\theta\in\Theta}{\sup}\quad \|\nabla_\theta \log f_i(\obs;\theta)\|^2\right]    \right\}\]

which is $o_p(1)$ du to the consistency of $\emvB\theta$ and assumption \eqref{ass:regularity 0-4}. Which concludes the proof of theorem \eqref{prop:LRTbootsasym not iid}.

\subsection{Proof of proposition \ref{prop:verif G} }

Recall that we want to show that : 

\[\quad\underset{\theta'\in\Theta}{\sup} \E_{\theta'}\{\underset{\theta\in\Theta}{\sup}\|\nabla^k_\theta\log f_i(\obs_i;\theta)\|^\gamma\}<+\infty \]

 with $\gamma=2$ for $k=0,3,4$ and $\gamma=3$ for $k=1,2$, for every $i\in\mathds{N}$.

For a sake of clarity, we consider the following simplified notations : 

$g(\cov_{ij},\beta,\Lambda\xi_i) = g^{ij}_\theta(\xi_i)$ and $ g_\theta^i(\xi_i) = ( g^{ij}_\theta(\xi_i))_{j=1,...,J_i}$

$f(\obs_i;\theta) = \E\{f(\obs_i;\xi,\theta)\} \propto \E[\exp\left\{-V(\theta,\obs_i,\xi_i)\right\}]$ with 
\[V(\theta,\obs_i,\xi_i) = \frac{\sum_j(\obs_{ij}-g^j_\theta(\xi_i))^2}{2\sigma^2} = \frac{\|\obs_{i}-g_\theta^i(\xi_i)\|^2}{2\sigma^2}\]
where the expectation is taken with respect to the random variable $\xi$. 

As these quantities are individual, we get rid of the subscript $i$ to lighten the notations.

As we consider the parameter space $\Theta$ compact, the residual variance $\sigma$ is restricted to lie in a segment $[\sigma_{min};\sigma_{max}]$ with $0<\sigma_{min}<\sigma_{max}<+\infty$. That is why we consider the gaussian density up to a constant that won't change the reasonning. From now on we won't write $\propto$ and we make the shortcut $\Fc\xi=e^{-V(\theta,\obs,\xi)}$.

We also write $\E^Z\{\cdot\}$ when the expectation is taken with respect to the random variable $Z$. 

We suppose now that $\|g_\theta(\xi)\|\rightarrow +\infty$  as $\|\xi\|\rightarrow +\infty$ which is the most complicated case.  

We first deal with the simplest case $k=0$, we want to show that 

\[\E_{\theta'}\left\{\underset{\theta\in\Theta}{\sup}|\log\F|^2\right\}<+\infty\]

Let $\theta,\theta'\in\Theta$, let $M>0$,

\begin{align*}
    \F&= \E^\xi\left\{\Fc\xi\right\}\\
    &\geq\E^\xi\left\{\Fc\xi \mathds{1}(\xi\leq M)\right\}\\
    &\geq \E^\xi\{e^{-V(\theta,\obs,\xi)}\mathds{1}(\xi\leq M)\}\\
    &= \E^\xi\{e^{-\frac{\|\obs-g_\theta(\xi)\|^2}{2\sigma}}\mathds{1}(\xi\leq M)\} \\
    & = \E^\xi\{e^{-\frac{\|\obs\|^2+\|g_\theta(\xi)\|^2 -2 \obs^Tg_\theta(\xi)}{2\sigma}}\mathds{1}(\xi\leq M)\} \\
    &\geq \E^\xi\{e^{-\frac{\|\obs\|^2+\|g_\theta(\xi)\|^2 +2 \|\obs\|\|g_\theta(\xi)\|}{2\sigma}}\mathds{1}(\xi\leq M)\} 
\end{align*}

where the last inequality is a direct application of Cauchy Schwartz's inequality. 

The quantity in the exponential is a polynomial in $\|g_\theta(\xi)\|$ that goes to $-\infty$ when $\|\xi\|\rightarrow +\infty$. Its minimal value is achieved at $\alpha_M(\theta)= \underset{\xi:\|\xi\|\leq M}{\sup}\quad \|g_\theta(\xi)\| = \|g_\theta(\tilde\xi)\|$ 

Therefore we obtain that for every $M>0$, 

\begin{equation}\label{eq:upper}
    \F  \geq \pr(\xi\leq M) e^{-\frac{(\|\obs\|+\alpha_M(\theta))^2}{2\sigma^2}}
\end{equation}

yet, by writing $\kappa_M =  \pr(\xi\leq M)$, 

\begin{align*}
    & \quad1>\F> \kappa_M e^{-\frac{(\|\obs\|+\alpha_M(\theta))^2}{2\sigma^2}}\\
    \Leftrightarrow &\quad 0> \log \F > \log(\kappa_M) - \frac{(\|\obs\|+\alpha_M(\theta))^2}{2\sigma^2} \\ 
    \Leftrightarrow &\quad 0 < |\log \F|^2 < \log(\kappa_M)^2 + \frac{(\|\obs\|+\alpha_M(\theta))^4}{4\sigma^2} - \frac{\log(\kappa_M)(\|y\|+\alpha_M(\theta))^2}{\sigma^2}\\ 
    \Leftrightarrow &\quad 0 < |\log \F|^2 < \log(\kappa_M)^2 + \frac{\{\|\obs\|+\alpha_M(\theta)\}^4}{4\sigma^4} \\
    \Leftrightarrow &\quad 0 < |\log \F|^2 < \log(\kappa_M)^2 + \frac{\{\|\obs\|+\alpha_M(\Bar{\theta})\}^4}{4\sigma_{min}^4}
\end{align*}

where $\Bar\theta = \underset{\theta\in\Theta}{\text{argsup}}\hspace{0.1cm}\alpha(\theta) \in \Theta$ by continuity of $\alpha(\cdot)$ (continuity of $\theta\rightarrow g_\theta(\tilde\xi)$) and compactness of $\Theta$

We define the quantity $P_M(\obs) =\log(\kappa_M)^2 + \frac{\{\|\obs\|+\alpha_M(\Bar\theta)\}^4}{4\sigma_{min}^4} $ that does not depend on $\theta$.

Therefore we have that : 

\begin{align*}
    \E_{\theta'}\left\{\underset{\theta\in\Theta}{\sup}\quad|\log \F|^2  \right\} & \leq  \E_{\theta'}\left\{ P_M(\obs) \right\} \\
    &= \int_\obs P_M(y)\Fp dy\\
    &= \int_\obs P_M(\obs) \int_\xi \Fcp\xi \pi_\df(\xi)d\xi d\obs\\
     & = \int_\xi\int_\obs  P_M(\obs) \Fcp\xi \pi_\df(\xi) d\obs d\xi\\
     &= \int_\xi\int_u 2\sigma'^2P_M(2\sigma'^2u+g_{\theta'}(\xi)) \pi_J(u)\pi_\df(\xi)dud\xi\\
    &\leq \int_\xi\int_u 2\sigma_{max}^2P_M(2\sigma_{max}^2u+g_{\theta'}(\xi)) \pi_J(u)\pi_\df(\xi)dud\xi
\end{align*}

using first Fubini-Tonelli's theorem and then a change of variable : $u = \frac{\obs-g_{\theta'}(\xi)}{2\sigma'^2}$.

$P_M(2\sigma'^2u+g_{\theta'}(\xi))$ is a polynomial of degree $4$ in $\|u\|$ and $\|g_{\theta'}(\xi)\|$. Thanks to the assumption \eqref{eq:reg_g} of proposition \eqref{prop:verif G} with $k_1=0$ and $k_2=4$, we have that :
\begin{align*}
    \underset{\theta'\in\Theta}{\sup}\E_{\theta'}\left\{\underset{\theta\in\Theta}{\sup}\quad|\log \F|^2  \right\}& \leq \int_\xi\int_u 2\sigma_{max}^2 \underset{\theta'\in\Theta}{\sup}P_M(2\sigma_{max}^2u+g_{\theta'}(\xi)) \Psi_J(u)\Psi_\df(\xi)dud\xi
\end{align*}

which concludes the first part of this proof. 

We now consider the case $k=1,2,3,4$, and $\gamma=2,3$ (the fact of considering $\gamma=2$ or $3$ is the same, therefore we will consider $\gamma=3$ as it is stronger). 

\begin{equation*}
    \|\nabla_\theta^k\log \F\|^3\leq \underset{I=(i_1,...,i_k)}{\underset{I\in\{1,...,d_\theta\}^k}{\sup} }d_\theta^k\|\frac{\partial^k \log \F}{ \prod_{j=1}^k\partial \theta_{i_j}}\|^3
\end{equation*}

Let $I_0=(i_1,...,i_k)$ the subset of indexes where the sup in the right hand side is achieved. \\ 

We consider the quantity 

\[\frac{\partial^k \log \F}{ \prod_{j=1}^k\partial \theta_{i_j}}\]

as the derivatives of the comoposition

\[\theta \mapsto \F \mapsto \log\F \]

and we use Faa di Bruno's formula to develop this expression : 

\[\frac{\partial^k \log \F}{ \prod_{j=1}^k\partial \theta_{i_j}} = \sum_{\Psi\in \mathcal{P}(\{i_1,...,i_k\})} \alpha_\Psi \F^{-|\Psi|} \prod_{B\in\Psi} \frac{\partial^{|B|}\F}{\prod_{b\in B}\partial \theta_b} \]

where $\mathcal{P}(K)$ stands for all partitions of a set $K$, and $ (\alpha_\Psi)_\Psi$ are constants. A sufficient condition for this quantity to be $\mathds{L}^3$ is that each term of the sum is $\mathds{L}^3$. \vspace{0.5cm}

Let $\Psi\in\mathcal{P}(\{i_1,...,i_k\})$, we write $m$ the cardinal of $\Psi$. 

We recall that :

\[\F = \E^\xi\left\{e^{-V(\theta,y,\xi)}\right\}\]

therefore, for every $B\in\Psi$,

\begin{align*}
    \frac{\partial^{|B|}\F}{\prod_{b\in B}\partial \theta_b} &= \E^\xi\left\{  \frac{\partial^{|B|}\Fc\xi}{\prod_{b\in B}\partial \theta_b} \right\} \\
    & = \E^\xi\left[P^B\{V(\theta,\obs,\xi)\}e^{-V(\theta,y,\xi)}\right]
\end{align*}

where $P^B\{V(\theta,\obs,\xi)\}$ is a polynomial of degree $m$ in $\obs$ and the partial derivatives of $g_\theta(\xi)$. 

\begin{align*}
    |\F^{-m} \prod_{B\in\Psi} \frac{\partial^{|B|}\F}{\prod_{b\in B}\partial \theta_b}| & = \F^{-m} \prod_{B\in\Psi} |\frac{\partial^{|B|}\F}{\prod_{b\in B}\partial \theta_b}| \\
    (Jensen)&\leq \F^{-m} \prod_{B\in\Psi} \E^\xi\left[|P^B\{V(\theta,\obs,\xi)\}|e^{-V(\theta,y,\xi)}\right]
\end{align*}

the random variables $\xi$ can be renamed $\xi_B$ for each term of the product so that : 

\[|\F^{-m} \prod_{B\in\Psi} \frac{\partial^{|B|}\F}{\prod_{b\in B}\partial \theta_b}|\leq  \F^{-m} \prod_{B\in\Psi} \E^{\xi_B}\left[|P^B\{V(\theta,\obs,\xi_B)\}|e^{-V(\theta,y,\xi_B)}\right]\]

We introduce $\Xi = (\xi_{B_1}^T, ...,\xi_{B_m}^T)^T\sim \mathcal{N}(0,\Id{m\times\df})$, so that we can write :

\begin{align*}
\F^{-m} \prod_{B\in\Psi} \E^{\xi_B}&\left[|P^B\{V(\theta,\obs,\xi_B)\}|e^{-V(\theta,y,\xi_B)}\right]\\ 
&= \F^{-m}  \E^{\Xi}\left[\prod_{B\in\Psi}|P^B\{V(\theta,\obs,\xi_B)\}|e^{-V(\theta,y,\xi_B)}\right] \\
&=  \F^{-m}  \E^{\Xi}\left(\left[\prod_{B\in\Psi}|P^B\{V(\theta,\obs,\xi_B)\}|\right]e^{-\sum_{B\in\Psi}V(\theta,y,\xi_B)}\right)
\end{align*}
every terms in the product inside the expectation is nonnegative, therefore when rising this quantity to the power of $3$ we can use Jensen by convexity on $\bbr^+$ of $x\mapsto x^3$. And we find that  : 

\begin{align*}
    |\F^{-m} \prod_{B\in\Psi} \frac{\partial^{|B|}\F}{\prod_{b\in B}\partial \theta_b}|^3 &\leq \F^{-3m}  \E^{\Xi}\left(\left[\prod_{B\in\Psi}|P^B\{V(\theta,\obs,\xi_B)\}|\right]e^{-\sum_{B\in\Psi}V(\theta,y,\xi_B)}\right)^3\\
    &\leq \F^{-3m}  \E^{\Xi}\left(\left[\prod_{B\in\Psi}|P^B\{V(\theta,\obs,\xi_B)\}|^3\right]e^{-\sum_{B\in\Psi}3V(\theta,y,\xi_B)}\right)
\end{align*}
using Jensen's inequality.\vspace{0.5cm}

By using equation \eqref{eq:upper}, we know that :

\[\F  \geq \kappa_M e^{-\frac{(\|\obs\|+\alpha_M(\Bar\theta))^2}{2\sigma^2}}\] 

$\kappa_m$ is a constant that has no impact on the reasoning therefore we neglect it to lighten the notations. We write $V(\obs,\sigma) =\frac{(\|\obs\|+\alpha_M(\Bar\theta))^2}{2\sigma^2} $ so that we have : 

\begin{align*}
    |\F^{-m} \prod_{B\in\Psi} \frac{\partial^{|B|}\F}{\prod_{b\in B}\partial \theta_b}|^3 \leq e^{3mV(\obs,\sigma)}\E^{\Xi}\left(\left[\prod_{B\in\Psi}|P^B\{V(\theta,\obs,\xi_B)\}|^3\right]e^{-\sum_{B\in\Psi}3V(\theta,y,\xi_B)}\right)
\end{align*}

We recall that the cardinal of $\Psi$ is equal to $m$ so : 

\begin{equation}\label{eq:intermediaire2}
    |\F^{-m} \prod_{B\in\Psi} \frac{\partial^{|B|}\F}{\prod_{b\in B}\partial \theta_b}|^3 \leq \E^{\Xi}\left(\left[\prod_{B\in\Psi}|P^B\{V(\theta,\obs,\xi_B)\}|^3\right]e^{-3\sum_{B\in\Psi}V(\theta,y,\xi_B) - V(\obs,\sigma)}\right)
\end{equation}

Once more to lighten the notations we define $M_\theta^\Psi( \obs,\Xi) = \prod_{B\in\Psi}|P^B\{V(\theta,\obs,\xi_B)\}|^3$\\ 

Let $\theta'\in\Theta$, 

\begin{align}\label{eq:intermediaire}
    &\E_{\theta'}^\obs\left\{|\F^{-m} \prod_{B\in\Psi} \frac{\partial^{|B|}\F}{\prod_{b\in B}\partial \theta_b}|^3 \right\}\nonumber\\
    &\leq  \E_{\theta'}^\obs\left\{\E^{\Xi}\left(M_\theta^\Psi( \obs,\Xi)e^{-3\sum_{B\in\Psi}V(\theta,y,\xi_B) - V(\obs,\sigma)}\right)\right\} \nonumber\\
    & =  \E^{\Xi}\left\{\E_{\theta'}^\obs\left(M_\theta^\Psi( \obs,\Xi)e^{-3\sum_{B\in\Psi}V(\theta,y,\xi_B) - V(\obs,\sigma)}\right)\right\}\nonumber\\
    &= \E^{\Xi}\left\{\int_\obs M_\theta^\Psi( \obs,\Xi)e^{-3\sum_{B\in\Psi}V(\theta,y,\xi_B) - V(\obs,\sigma)}\Fp d\obs\right\}\nonumber\\
    &= \E^{\Xi}\left[\int_\obs M_\theta^\Psi( \obs,\Xi)e^{-3\sum_{B\in\Psi}V(\theta,y,\xi_B) - V(\obs,\sigma)}\E^\xi\{e^{-V(\theta',\obs,\xi)}\} d\obs\right]\nonumber\\
     & =\E^{\Xi,\xi}\left\{\int_\obs M_\theta^\Psi( \obs,\Xi)e^{-3\sum_{B\in\Psi}V(\theta,y,\xi_B) - V(\obs,\sigma)}e^{-V(\theta',\obs,\xi)}d\obs\right\}
\end{align}
using once more Fubini Tonelli's theorem.\vspace{0.5cm}

We focus on the exponential : 

\begin{align*}
-3\sum_{B\in\Psi}\{V(\theta,y,\xi_B) - &V(\obs,\sigma)\} -V(\theta',\obs,\xi) =-3\sum_{B\in\Psi}\left\{V(\theta,y,\xi_B) - V(\obs,\sigma) \right\} - V(\theta',\obs,\xi)\\
&  = -3 \sum_{B\in\Psi} \left\{\frac{\|\obs-g_\theta(\xi_B)\|^2}{2\sigma^2} -\frac{\{\|\obs\|+\alpha(\Bar\theta)\}^2}{2\sigma^2} \right\}-\frac{\|\obs-g_{\theta'}(\xi)\|^2}{2\sigma'^2}\\
 &\begin{aligned}
= -3\sum_{B\in\Psi} \frac{1}{2\sigma^2}\{\|g_\theta(\xi_B)\|^2 & - 2\obs^Tg_\theta(\xi_B) + 2\|\obs\|\alpha(\Bar\theta)- \alpha(\bar\theta)^2\} \\&- \frac{1}{2\sigma'^2}\{ \|\obs\|^2 + \|g_{\theta'}(\xi)\|^2 -2 \obs^Tg_{\theta'}(\xi) \}\end{aligned}
\end{align*}

We use Cauchy-Schwartz's inequality to get rid of the scalar product and consider scalar quantities : 

\begin{align*}
-3\sum_{B\in\Psi}\{V(\theta,y,\xi_B) - &V(\obs,\sigma)\} -V(\theta',\obs,\xi) =-3\sum_{B\in\Psi}\left\{V(\theta,y,\xi_B) - V(\obs,\sigma) \right\} - V(\theta',\obs,\xi)\\
& \begin{aligned}
\leq -3\sum_{B\in\Psi}& \frac{1}{2\sigma^2}\{\|g_\theta(\xi_B)\|^2  - 2\|\obs\|\|g_\theta(\xi_B)\| + 2\|\obs\|\alpha(\Bar\theta) - \alpha(\bar\theta)^2\} \\
&- \frac{1}{2\sigma'^2}\{ \|\obs\|^2 + \|g_{\theta'}(\xi)\|^2 -2 \|\obs\|\|g_{\theta'}(\xi)\| \}
\end{aligned}
\end{align*}

Therefore by taking the integrated form of \eqref{eq:intermediaire} we have that : 

\begin{equation*}
  \begin{split}
       \E_{\theta'}^\obs&\left\{|\F^{-m} \prod_{B\in\Psi} \frac{\partial^{|B|}\F}{\prod_{b\in B}\partial \theta_b}|^3 \right\} \\&\leq \int_{\Xi,\xi}\int_\obs M_\theta^\Psi( \obs,\Xi)e^{-3\sum_{B\in\Psi}V(\theta,y,\xi_B) - V(\obs,\sigma)}e^{-V(\theta',\obs,\xi)}d\obs e^{-\frac{\|\Xi\|^2}{2} - \frac{\|\xi\|^2}{2}} d\Xi d\xi\\
       &\leq \int_{\Xi,\xi,\obs}M_\theta^\Psi( \obs,\Xi)e^{-3\sum_{B\in\Psi}V(\theta,y,\xi_B) - V(\obs,\sigma)-V(\theta',\obs,\xi)-\frac{\|\Xi\|^2}{2} - \frac{\|\xi\|^2}{2} } d\obs d\Xi d\xi\\
       &\leq \int_{\Xi,\xi,\obs}M_\theta^\Psi( \obs,\Xi)e^{H(\theta, \theta', \obs, \Xi, \xi) } d\obs d\Xi d\xi\\
\end{split}
\end{equation*}

where 
\begin{equation*}
    \begin{split}
        H(\theta, \theta', \obs, \Xi, \xi) = -3\sum_{B\in\Psi}\frac{1}{2\sigma^2}\{\|g_\theta(\xi_B)\|^2  - 2\|\obs\|\|g_\theta(\xi_B)\| + 2\|\obs\|\alpha(\Bar\theta)- \alpha(\bar\theta)^2\} \\
- \frac{1}{2\sigma'^2}\{ \|\obs\|^2 + \|g_{\theta'}(\xi)\|^2 -2 \|\obs\|\|g_{\theta'}(\xi)\| -\frac{\|\Xi\|^2}{2} - \frac{\|\xi\|^2}{2}
    \end{split}
\end{equation*}

by splitting $\|\Xi\|^2$ and rearranging the terms we get : 

\begin{align*}
    H(\theta, \theta', \obs, \Xi, \xi) = \|\obs\|\left\{\frac{\|g_{\theta'}(\xi)\| }{\sigma'^2}- \frac{3m}{\sigma^2}\alpha(\Bar\theta)  \right\}+\frac{3m}{2\sigma^2}\alpha(\Bar\theta)^2 +\sum_{B\in\Psi}\frac{3}{\sigma^2}   \|\obs\|\|g_\theta(\xi_B)\|  \\
- \frac{1}{2\sigma'^2} \|\obs\|^2    - \frac{1}{2}\left\{\|\xi\|^2 +  \frac{1}{\sigma'^2}\|g_{\theta'}(\xi)\|^2\right\} -\frac{1}{2}\sum_{B\in\Psi}\left\{\frac{3}{\sigma^2}\|g_\theta(\xi_B)\|^2 +\|\xi_B\|^2\right\}
\end{align*}

As it is constant we can omit $\frac{3m}{2\sigma^2}\alpha(\Bar\theta)^2$ in the development as it can be taken out from the integrals. Furhtermore $-\alpha(\Bar\theta\|\obs\|/\sigma^2<0$ therefore it can be upper bounded by zero. 

Then we define for $\theta\in\Theta$, $\xi\in\bbr^\df$ $r_\theta(\xi)$ the ratio $\frac{\|g_\theta(\xi)\|}{\|\xi\|}$ such that,

\begin{align*}
 H(\theta, \theta', \obs, \Xi, \xi) \leq &\frac{r_{\theta'}(\xi) }{\sigma'^2 }\|\xi\|  \|\obs\| + \sum_{B\in\Psi}\frac{3r_\theta(\xi_B) }{\sigma^2}  \|\xi_B\|\|\obs\|\\
  & - \frac{1}{2\sigma'^2} \|\obs\|^2    - \frac{1}{2}\left(1+\frac{r_{\theta'}(\xi)^2}{\sigma'^2}\right)\|\xi\|^2 - \frac{1}{2}\sum_{B\in\Psi}\left\{1+\frac{3r_{\theta}(\xi_B)^2}{\sigma^2}\right\}\|\xi_B\|^2 \\ 
  & \leq -\frac{1}{2} (\frac{\obs^T}{\sigma_{min}}, \tilde\xi^T,\tilde\Xi^T)^T \begin{pmatrix}
      1 & - \frac{\sigma_{min}r_{\theta'}(\xi)}{\sigma_{max}^2} & ... & -\frac{3\sigma_{min}r_\theta(\xi_B)}{\sigma_{max}^2}& ... \\
      - \frac{\sigma_{min}r_{\theta'}(\xi)}{\sigma_{max}^2} &  1 & 0 & \cdots & 0 \\
      \vdots &0 & \ddots &0 & \vdots \\
      -\frac{3\sigma_{min}r_\theta(\xi_B)}{\sigma_{max}^2} & \vdots & 0&\ddots & 0\\
      \vdots & 0 & \cdots & 0& 1
  \end{pmatrix}\begin{pmatrix}
      \frac{\obs}{\sigma_{min}} \\ \tilde\xi \\ \tilde\Xi 
  \end{pmatrix}
    \end{align*}
Where $\tilde\xi = \sqrt{\left(1+\frac{r_{\theta'}(\xi)^2}{\sigma'^2}\right)}\times\xi$ and $\tilde\Xi = \left(\sqrt{\left\{1+\frac{3r_{\theta}(\xi_B)^2}{\sigma^2}\right\}}\times\xi_B\right)_{B\in\Psi} $
Let $\Omega(\theta,\theta', \obs,\xi,\Xi)$ the symmetric matrix involved in the previous equation. In order to bound this last quantity we study the spectrum of this matrix. To do so we determine its characteristic.   Let $x\in\bbr$. We call $n= n$ By using the cofactor expansion formula for the first line we find  :  

\begin{align*}
    det\left\{x\Id{n}-\Omega(\theta,\theta', \obs,\xi,\Xi)\right\} &= (x-1)^{n} - \frac{\sigma_{min}^2r^2_{\theta'}(\xi)}{\sigma_{max}^4}(x-1)^{n-2} -\sum_{B\in\Psi} \frac{9\sigma^2_{min}r^2_\theta(\xi_B)}{\sigma_{max}^4}(x-1)^{n-2} \\
    &= (x-1)^{n-2}\left[ (x-1)^2 - \left\{\frac{\sigma_{min}^2r^2_{\theta'}(\xi)}{\sigma_{max}^4} +\sum_{B\in\Psi} \frac{9\sigma^2_{min}r^2_\theta(\xi_B)}{\sigma_{max}^4} \right\}\right]
\end{align*}

therefore the spectrum of $\Omega(\theta,\theta', \obs,\xi,\Xi)$ is 
\[\left\{ 1 ; 1 - \sqrt{\frac{\sigma_{min}^2r^2_{\theta'}(\xi)}{\sigma_{max}^4} +\sum_{B\in\Psi} \frac{9\sigma^2_{min}r^2_\theta(\xi_B)}{\sigma_{max}^4} };  1 +\sqrt{\frac{\sigma_{min}^2r^2_{\theta'}(\xi)}{\sigma_{max}^4} +\sum_{B\in\Psi} \frac{9\sigma^2_{min}r^2_\theta(\xi_B)}{\sigma_{max}^4} }  \right\}\] 

and we get : 
\begin{align*}
     H(\theta, \theta', \obs, \Xi, \xi) \leq& -\frac{1}{2}\left\{1 - \sqrt{\frac{\sigma_{min}^2r^2_{\theta'}(\xi)}{\sigma_{max}^4} +\sum_{B\in\Psi} \frac{9\sigma^2_{min}r^2_\theta(\xi_B)}{\sigma_{max}^4} }\right\}\left(\frac{\|\obs\|^2}{\sigma_{min}^2} + \|\tilde\xi\|^2 + \|\tilde\Xi\|^2\right)\\ 
     &\leq -\frac{1}{2}\left\{1 - \sqrt{\frac{\sigma_{min}^2r^2_{\theta'}(\xi)}{\sigma_{max}^4} +\sum_{B\in\Psi} \frac{9\sigma^2_{min}r^2_\theta(\xi_B)}{\sigma_{max}^4} }\right\}\left(\frac{\|\obs\|^2}{\sigma_{min}^2} + \|\xi\|^2 + \|\Xi\|^2\right)
\end{align*}

where the last equality holds because $\|\tilde\xi\|>\|\xi\|$ and $\|\tilde\Xi\|>\|\Xi\|$.\\

Let $\varepsilon >0$ small enough such that : 

\[ 0< 1 - \sqrt{\frac{\sigma_{min}^2\varepsilon^2}{\sigma_{max}^4} +\sum_{B\in\Psi} \frac{9\sigma^2_{min}\varepsilon^2}{\sigma_{max}^4}}<1  \]

Let $\omega = \sqrt{\frac{\sigma_{min}^2\varepsilon^2}{\sigma_{max}^4} +\sum_{B\in\Psi} \frac{9\sigma^2_{min}\varepsilon^2}{\sigma_{max}^4}}$

Let $K$ a compact set such that assumption \eqref{eq:reg_g2} is verified, and $K_M = \{\obs\in\bbr^J\leq M\}$ for some $M>0$

we finally get to : 

\begin{align*}
\begin{split}
    \E_{\theta'}^\obs&\left\{|\F^{-m} \prod_{B\in\Psi} \frac{\partial^{|B|}\F}{\prod_{b\in B}\partial \theta_b}|^3 \right\} \\&\leq \int_{(\Xi,\xi,\obs)\in K^{m+1}\times K_M}M_\theta^\Psi( \obs,\Xi)e^{H(\theta, \theta', \obs, \Xi, \xi)} d\obs d\Xi d\xi\\ & + \int_{(\Xi,\xi,\obs)\in \bbr^{(m+1)\df+J}\backslash K^{m+1}\times K_M}M_\theta^\Psi( \obs,\Xi)e^{-\frac{1}{2}(1-\omega)\left(\frac{\|\obs\|^2}{\sigma_{min}^2} + \|\xi\|^2 + \|\Xi\|^2\right) } d\obs d\Xi d\xi
    \end{split}
\end{align*}

Now, to be precise, we restart from equation \eqref{eq:intermediaire2} and the following calculation leading to equation \eqref{eq:intermediaire}, to show that the sup can be considered inside the integral :  

\begin{align*}
    &\E_{\theta'}^\obs\left\{\underset{\theta\in\Theta}{\sup}\quad|\F^{-m} \prod_{B\in\Psi} \frac{\partial^{|B|}\F}{\prod_{b\in B}\partial \theta_b}|^3 \right\}\\
    &\leq  \E_{\theta'}^\obs\left\{\underset{\theta\in\Theta}{\sup}\quad\E^{\Xi}\left(M_\theta^\Psi( \obs,\Xi)e^{-3\sum_{B\in\Psi}V(\theta,y,\xi_B) - V(\obs,\sigma)}\right)\right\} \\ 
    &= \int_\obs\underset{\theta\in\Theta}{\sup}\left[\E^{\Xi}\left\{M_\theta^\Psi( \obs,\Xi)e^{-3\sum_{B\in\Psi}V(\theta,y,\xi_B) - V(\obs,\sigma)}\right\}\right]f_{\theta'}(\obs)d\obs \\ 
   (f_\theta'(\obs)>0) &= \int_\obs\underset{\theta\in\Theta}{\sup}\left[\E^{\Xi}\left\{M_\theta^\Psi( \obs,\Xi)e^{-3\sum_{B\in\Psi}V(\theta,y,\xi_B) - V(\obs,\sigma)}\right\}f_{\theta'}(\obs)\right]d\obs\\
    (Jensen)&\leq \int_\obs\E^{\Xi}\left[\underset{\theta\in\Theta}{\sup}\left\{M_\theta^\Psi( \obs,\Xi)e^{-3\sum_{B\in\Psi}V(\theta,y,\xi_B) - V(\obs,\sigma)}\right\}\right]f_{\theta'}(\obs)d\obs
\end{align*}

where the second equality holds because $f_{\theta'}(\obs)$ is nonnegative and does not depend on $\theta$. And finally : 

\begin{align}\label{eq:final}
\begin{split}
    \E_{\theta'}^\obs&\left\{\underset{\theta\in\Theta}{\sup}|\F^{-m} \prod_{B\in\Psi} \frac{\partial^{|B|}\F}{\prod_{b\in B}\partial \theta_b}|^3 \right\} \\&\leq \int_{(\Xi,\xi,\obs)\in K^{m+1}\times K_M}\underset{\theta\in\Theta}{\sup }M_\theta^\Psi( \obs,\Xi)e^{H(\theta, \theta', \obs, \Xi, \xi)} d\obs d\Xi d\xi\\ & + \int_{(\Xi,\xi,\obs)\in \bbr^{(m+1)\df+J}\backslash K^{m+1}\times K_M}\underset{\theta\in\Theta}{\sup} M_\theta^\Psi( \obs,\Xi)e^{-\frac{1}{2}(1-\omega)\left(\frac{\|\obs\|^2}{\sigma_{min}^2} + \|\xi\|^2 + \|\Xi\|^2\right) } d\obs d\Xi d\xi \\
    &\leq \underset{(\Xi,\xi,\obs)\in K^{m+1}\times K_M}{\underset{\theta\in\Theta }{\sup }}Vol(K^{m+1}\times K_M) M_\theta^\Psi( \obs,\Xi)e^{H(\theta, \theta', \obs, \Xi, \xi)} \\ & + \int_{(\Xi,\xi,\obs)\in \bbr^{(m+1)\df+J}\backslash K^{m+1}\times K_M}\underset{\theta\in\Theta}{\sup} M_\theta^\Psi( \obs,\Xi)e^{-\frac{1}{2}(1-\omega)\left(\frac{\|\obs\|^2}{\sigma_{min}^2} + \|\xi\|^2 + \|\Xi\|^2\right) } d\obs d\Xi d\xi
    \end{split}
    \end{align}
The first term of the last quantity is a suprema of a continuous function over a compact set and is therefore finite.

$M_\theta^\Psi(\obs,\Xi)$ is a polynomial with respect to $\|\obs\|$ and the partial derivatives of $\theta\mapsto g_\theta(\xi_B)$ for each $B\in \Psi$, therefore, thanks to assumption \eqref{eq:reg_g}, considering $\varepsilon$ small enough such that $\varepsilon<\delta$ where $\delta$ is defined in proposition \eqref{prop:verif G}, we finally get to the conclusion : 

\[\E_{\theta'}^\obs\left\{\underset{\theta\in\Theta}{\sup}|\F^{-m} \prod_{B\in\Psi} \frac{\partial^{|B|}\F}{\prod_{b\in B}\partial \theta_b}|^3 \right\}<+\infty\]

In equation \eqref{eq:final} is is important to notice that the last term no longer depends on $\theta'$, and that the first term in the sup is a continuous function of $\theta'$, therefore by compacity of $\Theta$ the suprema can be taken with respect to $\theta'$ and

\[\underset{\theta'\in\Theta}{\sup}\quad\E_{\theta'}^\obs\left\{\underset{\theta\in\Theta}{\sup}|\F^{-m} \prod_{B\in\Psi} \frac{\partial^{|B|}\F}{\prod_{b\in B}\partial \theta_b}|^3 \right\}<+\infty\]
 which concludes the proof. 

\subsection{Linear model specific case}\label{sec:linear}
when considering a linear model, for every $i=1,...,N$  the model writes : 

$$y_{ij} = x_{i}^T\beta + Z_{i}\Lambda\xi_i +\varepsilon_{i} $$
where $\xi_i\sim\mathcal{N}\left(0,I\right)$ and independent from $\varepsilon_{i}\sim\mathcal{N}(0,\sigma^2I_{J_i})$. Therefore we have :

\[y_{ij} \sim \mathcal{N}\left(  x_{i}^T\beta , Z_{i}\Lambda\Lambda^TZ_i^T+\sigma^2I_{J_i}   \right)\]

And the log-likelihood is explicit : 

{\footnotesize\[\log f(y_i;\theta) = \frac{-J_i}{2}\log det\left(Z_{i}\Lambda\Lambda^TZ_i^T+\sigma^2I_{J_i} \right) - (y_i-X_i\beta)^T \left\{ Z_{i}\Lambda\Lambda^TZ_i^T+\sigma^2I_{J_i}\right\}^{-1}(y_i-X_i\beta) \]}

$y_i$ is Gaussian, therefore have all its moments finite. Given that $\sigma^2$ is strictly nonnegative, $\log f(y_i;\theta)$ is infinitely differentiable on $\Theta$, and each of its derivatives are quadratic forms of $y_i$, and therefore have finite moments. Which verifies assumption \eqref{ass:regularity 0-4}.
\subsection{Logistic growth model example}\label{logistic_validity}

We consider here the logistic growth model that is commonly used in many fields. We show how to use the criteria given in proposition \eqref{prop:verif G}. 

We recall the definition of the logistic growth function : 

\begin{equation}\label{eq:logisticmodel2}
    g(x_{ij},\beta,\Lambda \xi_i) = \frac{\beta_1+\lambda_1\xi_{i1}}{1 + \exp \left( -\frac{x_{ij}-(\beta_2+\lambda_2\xi_{i2})}{\beta_3+\lambda_3\xi_{i3}} \right)}.
\end{equation}

To verify that assumption \eqref{eq:reg_g} is verified, we need to calculate the derivatives of $g$ with respect to $\theta$. To avoid heavy calculations we consider the parameter $\lambda_3$ : 

\begin{align*}
    \frac{\partial g(x_{ij},\beta,\Lambda \xi_i)}{\partial \lambda_3} = &-\xi_{i3}\left( \frac{j-(\beta_2+\lambda_2\xi_{i2})}{(\beta_3+\lambda_3\xi_{i3})^2} \right) \exp \left( -\frac{j-(\beta_2+\lambda_2\xi_{i2})}{\beta_3+\lambda_3\xi_{i3}} \right) \\&\times \frac{\beta_1+\lambda_1\xi_{i1}}{\left(1 + \exp \left( -\frac{j-(\beta_2+\lambda_2\xi_{i2})}{\beta_3+\lambda_3\xi_{i3}} \right)\right)^2}
\end{align*}
The only issue of this quantity occurs at $\xi_{i3} = -\frac{\beta_3}{\lambda_3}$ that tends toward $0$ as $\xi_{i3} \rightarrow -\frac{\beta_3}{\lambda_3}$. Finally this quantity is integrable with respect to the Gaussian distribution. By iterating the derivatives, we find expressions similar to this one and assumption \eqref{eq:reg_g} is verified. 

Let $\varepsilon>0,M>0$, and we define the set  : 

\[K_M = \{\xi \in \bbr^3 : |\xi_1|\leq M, |\xi_2|\leq M |\xi_1|, |\xi_3|\leq M |\xi_1|\}\]

Therefore for every $\xi\in\bbr^3\backslash K_M$ : 
\begin{align*}
\frac{\|g(\cov_{ij},\beta,\Lambda\xi)\|^2}{\|\xi\|^2} &\leq \frac{\lambda_1^2\xi_1^2}{\xi_1^2+\xi_2^2+\xi_3^2}\\
&\leq \frac{\lambda_1^2\xi_1^2}{M^2\xi_1^2}\\
&\leq \frac{\lambda_1^2}{M^2}
\end{align*}

By taking $M>\frac{\lambda_1}{\varepsilon}$, it verifies assumption \eqref{eq:reg_g2}. 

\subsection{Pharmacokinetic model}\label{pharmaco}

The pharmokinetic study of theophylline concentration along time is a well known example in the literature of mixed effects models , see for example \cite{davidian2017nonlinear}. The simplified model is defined as follows, the theophylline concentration of patient $i$ ($i=1,...,N)$ at time $t_j$ ($j=1,...,J$) is modeled as : 

\[\left\{\begin{array}{c}
     y_{ij} = \frac{D{k_a}_i}{V_i{k_a}_i-Cl_i}\left\{\exp\left({-{k_a}_it_j}\right)-\exp\left({-\frac{Cl_i}{V_i}t_j}\right)\right\} +\varepsilon_{ij}, \quad \varepsilon_{ij}\sim\mathcal{N}(0,\sigma^2)  \\
     
      ({k_a}_i,Cl_i,V_i) = \left(\exp\left\{\beta_1+\lambda_1\xi_{i1}\right\},\exp\left\{\beta_2+\lambda_2\xi_{i2}\right\},\exp\left\{\beta_3+\lambda_3\xi_{i3}\right\}\right) \\
      (\xi_{i1},\xi_{i2},\xi_{i3})^T \sim\mathcal{N}\left(0,I_3\right)

\end{array}\right.\]

where $D$ is a specific constant of the experiment, and $({k_a}_i,Cl_i,V_i)$ are individual parameters describing the biological process (rate of drug absorption,  rate of drug elimination, complete volume of blood with drugs). The exponentials enforce positivity of these parameters. In this setting, 

\[g(x_{ij},\beta,\Lambda \xi_i) =\frac{D{k_a}_i}{V_i{k_a}_i-Cl_i}\left\{\exp\left({-{k_a}_it_j}\right)-\exp\left({-\frac{Cl_i}{V_i}t_j}\right)\right\}\]
(where  $x_{ij} = t_j$ and $\Lambda $ is diagonal with diagonal $(\lambda_1,\lambda_2,\lambda_3)^T$) is a bounded function of $(\xi_{i1},\xi_{i2},\xi_{i3})$, and therefore the criterion is straightforward to verify. 

\bibliographystyle{plainnat}
\bibliography{biblio}

\begin{thebibliography}{46}
\providecommand{\natexlab}[1]{#1}
\providecommand{\url}[1]{\texttt{#1}}
\expandafter\ifx\csname urlstyle\endcsname\relax
  \providecommand{\doi}[1]{doi: #1}\else
  \providecommand{\doi}{doi: \begingroup \urlstyle{rm}\Url}\fi

\bibitem[Andrews(1993)]{andrews1993tests}
Donald~WK Andrews.
\newblock Tests for parameter instability and structural change with unknown change point.
\newblock \emph{Econometrica: Journal of the Econometric Society}, pages 821--856, 1993.

\bibitem[Andrews(1999)]{andrews1999estimation}
Donald~WK Andrews.
\newblock Estimation when a parameter is on a boundary.
\newblock \emph{Econometrica}, 67\penalty0 (6):\penalty0 1341--1383, 1999.

\bibitem[Andrews(2000)]{andrews2000inconsistency}
Donald~WK Andrews.
\newblock Inconsistency of the bootstrap when a parameter is on the boundary of the parameter space.
\newblock \emph{Econometrica}, pages 399--405, 2000.

\bibitem[Baey et~al.(2019)Baey, Courn{\`e}de, and Kuhn]{baey2019asymptotic}
Charlotte Baey, Paul-Henry Courn{\`e}de, and Estelle Kuhn.
\newblock Asymptotic distribution of likelihood ratio test statistics for variance components in nonlinear mixed effects models.
\newblock \emph{Computational Statistics \& Data Analysis}, 135:\penalty0 107--122, 2019.

\bibitem[Beran(1997)]{beran1997diagnosing}
Rudolf Beran.
\newblock Diagnosing bootstrap success.
\newblock \emph{Annals of the Institute of Statistical Mathematics}, 49\penalty0 (1):\penalty0 1--24, 1997.

\bibitem[Bickel and Sakov(2008)]{bickel2008choice}
Peter~J Bickel and Anat Sakov.
\newblock On the choice of m in the m out of n bootstrap and confidence bounds for extrema.
\newblock \emph{Statistica Sinica}, pages 967--985, 2008.

\bibitem[Bolker et~al.(2009)Bolker, Brooks, Clark, Geange, Poulsen, Stevens, and White]{bolker2009}
Benjamin~M Bolker, Mollie~E Brooks, Connie~J Clark, Shane~W Geange, John~R Poulsen, M~Henry~H Stevens, and Jada-Simone~S White.
\newblock Generalized linear mixed models: a practical guide for ecology and evolution.
\newblock \emph{Trends in ecology \& evolution}, 24\penalty0 (3):\penalty0 127--135, 2009.

\bibitem[Bonate(2011)]{bonate2011}
Peter~L Bonate.
\newblock Nonlinear mixed effects models: theory.
\newblock \emph{Pharmacokinetic-pharmacodynamic modeling and simulation}, pages 233--301, 2011.

\bibitem[Brown and Prescott(2015)]{brown2015}
Helen Brown and Robin Prescott.
\newblock \emph{Applied mixed models in medicine}.
\newblock John Wiley \& Sons, 2015.

\bibitem[Bulinski(2017)]{bulinski2017conditional}
Alexander~V Bulinski.
\newblock Conditional central limit theorem.
\newblock \emph{Theory of Probability \& Its Applications}, 61\penalty0 (4):\penalty0 613--631, 2017.

\bibitem[Cavaliere et~al.(2020)Cavaliere, Nielsen, Pedersen, and Rahbek]{cavaliere2020bootstrap}
Giuseppe Cavaliere, Heino~Bohn Nielsen, Rasmus~S{\o}ndergaard Pedersen, and Anders Rahbek.
\newblock Bootstrap inference on the boundary of the parameter space, with application to conditional volatility models.
\newblock \emph{Journal of Econometrics}, 2020.

\bibitem[Chant(1974)]{Chant74}
D~Chant.
\newblock {On Asymptotic Tests of Composite Hypotheses in Nonstandard Conditions}.
\newblock \emph{Biometrika}, 61\penalty0 (2):\penalty0 291--298, 1974.

\bibitem[Chen and Dunson(2003)]{chen2003random}
Zhen Chen and David~B Dunson.
\newblock Random effects selection in linear mixed models.
\newblock \emph{Biometrics}, 59\penalty0 (4):\penalty0 762--769, 2003.

\bibitem[Chernoff(1954)]{Cher54}
Herman Chernoff.
\newblock {On the Distribution of the Likelihood Ratio}.
\newblock \emph{The Annals of Mathematical Statistics}, 25\penalty0 (3):\penalty0 573--578, 1954.

\bibitem[Crainiceanu and Ruppert(2004)]{Crai04}
Ciprian~M. Crainiceanu and David Ruppert.
\newblock {Likelihood ratio tests for goodness-of-fit of a nonlinear regression model}.
\newblock \emph{Journal of Multivariate Analysis}, 91\penalty0 (1):\penalty0 35--52, oct 2004.

\bibitem[Davidian and Giltinan(2017)]{davidian2017nonlinear}
Marie Davidian and David~M Giltinan.
\newblock \emph{Nonlinear models for repeated measurement data}.
\newblock Routledge, 2017.

\bibitem[Delattre et~al.(2014)Delattre, Lavielle, and Poursat]{delattre2014note}
Maud Delattre, Marc Lavielle, and Marie-Anne Poursat.
\newblock {A note on BIC in mixed-effects models}.
\newblock \emph{Electronic Journal of Statistics}, 8\penalty0 (1):\penalty0 456 -- 475, 2014.
\newblock \doi{10.1214/14-EJS890}.
\newblock URL \url{https://doi.org/10.1214/14-EJS890}.

\bibitem[Drikvandi et~al.(2013)Drikvandi, Verbeke, Khodadadi, and {Partovi Nia}]{Drik13}
R.~Drikvandi, G.~Verbeke, A.~Khodadadi, and V.~{Partovi Nia}.
\newblock {Testing multiple variance components in linear mixed-effects models}.
\newblock \emph{Biostatistics}, 14\penalty0 (1):\penalty0 144--159, 2013.

\bibitem[Ekvall and Bottai(2021)]{ekvall2021confidence}
Karl~Oskar Ekvall and Matteo Bottai.
\newblock Confidence regions near singular information and boundary points with applications to mixed models.
\newblock \emph{arXiv preprint arXiv:2103.10236}, 2021.

\bibitem[Geyer(1994)]{geyer1994asymptotics}
Charles~J Geyer.
\newblock On the asymptotics of constrained m-estimation.
\newblock \emph{The Annals of statistics}, pages 1993--2010, 1994.

\bibitem[Gin{\'e} and Nickl(2021)]{gine2021mathematical}
Evarist Gin{\'e} and Richard Nickl.
\newblock \emph{Mathematical foundations of infinite-dimensional statistical models}.
\newblock Cambridge university press, 2021.

\bibitem[Gordon(2019)]{gordon2019}
Katherine~R Gordon.
\newblock How mixed-effects modeling can advance our understanding of learning and memory and improve clinical and educational practice.
\newblock \emph{Journal of Speech, Language, and Hearing Research}, 62\penalty0 (3):\penalty0 507--524, 2019.

\bibitem[Goymann et~al.(2016)Goymann, Safari, Muck, and Schwabl]{goymann2016sex}
Wolfgang Goymann, Ignas Safari, Christina Muck, and Ingrid Schwabl.
\newblock Sex roles, parental care and offspring growth in two contrasting coucal species.
\newblock \emph{Royal Society Open Science}, 3\penalty0 (10):\penalty0 160463, 2016.

\bibitem[Groll and Tutz(2014)]{groll2014variable}
Andreas Groll and Gerhard Tutz.
\newblock Variable selection for generalized linear mixed models by l 1-penalized estimation.
\newblock \emph{Statistics and Computing}, 24:\penalty0 137--154, 2014.

\bibitem[Gurka(2006)]{gurka2006selecting}
Matthew~J Gurka.
\newblock Selecting the best linear mixed model under reml.
\newblock \emph{The American Statistician}, 60\penalty0 (1):\penalty0 19--26, 2006.

\bibitem[Hansen(2022)]{hansen2022econometrics}
Bruce Hansen.
\newblock \emph{Econometrics}.
\newblock Princeton University Press, 2022.

\bibitem[Higham(1990)]{higham1990analysis}
Nicholas~J Higham.
\newblock Analysis of the cholesky decomposition of a semi-definite matrix.
\newblock \emph{{Reliable Numerical Commputation}}, 1990.

\bibitem[Hiroyuki et~al.(2012)Hiroyuki, Katsumi, et~al.]{testinghiroyuki2012}
Kasahara Hiroyuki, Shimotsu Katsumi, et~al.
\newblock Testing the number of components in finite mixture models.
\newblock Technical report, Institute of Economic Research, Hitotsubashi University, 2012.

\bibitem[Hoadley(1971)]{hoadley1971asymptotic}
Bruce Hoadley.
\newblock Asymptotic properties of maximum likelihood estimators for the independent not identically distributed case.
\newblock \emph{The Annals of mathematical statistics}, pages 1977--1991, 1971.

\bibitem[Ibrahim et~al.(2011)Ibrahim, Zhu, Garcia, and Guo]{ibrahim2011fixed}
Joseph~G Ibrahim, Hongtu Zhu, Ramon~I Garcia, and Ruixin Guo.
\newblock Fixed and random effects selection in mixed effects models.
\newblock \emph{Biometrics}, 67\penalty0 (2):\penalty0 495--503, 2011.

\bibitem[Kingma and Welling(2020)]{kingma2014}
Diederik~P Kingma and Max Welling.
\newblock Auto-encoding variational bayes, 2020.
\newblock URL \url{https://arxiv.org/abs/1312.6114}.

\bibitem[Lin(1997)]{lin1997variance}
Xihong Lin.
\newblock Variance component testing in generalised linear models with random effects.
\newblock \emph{Biometrika}, 84\penalty0 (2):\penalty0 309--326, 1997.

\bibitem[Meteyard and Davies(2020)]{meteyard2020}
Lotte Meteyard and Robert~A.I. Davies.
\newblock Best practice guidance for linear mixed-effects models in psychological science.
\newblock \emph{Journal of Memory and Language}, 112:\penalty0 104092, 2020.

\bibitem[Moran(1971)]{moran1971uniform}
PAP Moran.
\newblock The uniform consistency of maximum-likelihood estimators.
\newblock \emph{Mathematical Proceedings of the Cambridge Philosophical Society}, 70\penalty0 (3):\penalty0 435--439, 1971.

\bibitem[Nie(2006)]{nie2006strong}
Lei Nie.
\newblock Strong consistency of the maximum likelihood estimator in generalized linear and nonlinear mixed-effects models.
\newblock \emph{Metrika}, 63\penalty0 (2):\penalty0 123--143, 2006.

\bibitem[Pinheiro and Bates(2006)]{pinheiro2006mixed}
Jos{\'e} Pinheiro and Douglas Bates.
\newblock \emph{Mixed-effects models in S and S-PLUS}.
\newblock Springer science \& business media, 2006.

\bibitem[Rotnitzky et~al.(2000)Rotnitzky, Cox, Bottai, and Robins]{rotnitzky2000likelihood}
Andrea Rotnitzky, David~R Cox, Matteo Bottai, and James Robins.
\newblock Likelihood-based inference with singular information matrix.
\newblock \emph{Bernoulli}, pages 243--284, 2000.

\bibitem[Self and Liang(1987)]{self1987asymptotic}
Steven~G Self and Kung-Yee Liang.
\newblock Asymptotic properties of maximum likelihood estimators and likelihood ratio tests under nonstandard conditions.
\newblock \emph{Journal of the American Statistical Association}, 82\penalty0 (398):\penalty0 605--610, 1987.

\bibitem[Silvapulle and Sen(2005)]{silvapulle2005constrained}
Mervyn~J Silvapulle and Pranab~Kumar Sen.
\newblock \emph{Constrained statistical inference: Inequality, order and shape restrictions}.
\newblock John Wiley \& Sons, 2005.

\bibitem[Sinha(2009)]{sinha2009bootstrap}
Sanjoy~K Sinha.
\newblock Bootstrap tests for variance components in generalized linear mixed models.
\newblock \emph{Canadian Journal of Statistics}, 37\penalty0 (2):\penalty0 219--234, 2009.

\bibitem[Stram and Lee(1994)]{stramlee94}
DO~O Stram and JW~W Lee.
\newblock {Variance components testing in the longitudinal mixed effects model.}
\newblock \emph{Biometrics}, 50\penalty0 (4):\penalty0 1171--1177, 1994.

\bibitem[Vaida and Blanchard(2005)]{vaida2005conditional}
Florin Vaida and Suzette Blanchard.
\newblock Conditional akaike information for mixed effects models.
\newblock \emph{Corrado Lagazio, Marco Marchi (Eds)}, page 101, 2005.

\bibitem[Van~der Vaart(2000)]{van2000asymptotic}
Aad~W Van~der Vaart.
\newblock \emph{Asymptotic statistics}, volume~3.
\newblock Cambridge university press, 2000.

\bibitem[Wilks(1938)]{wilks}
S.~S. Wilks.
\newblock {The Large-Sample Distribution of the Likelihood Ratio for Testing Composite Hypotheses}.
\newblock \emph{The Annals of Mathematical Statistics}, 9\penalty0 (1):\penalty0 60 -- 62, 1938.
\newblock \doi{10.1214/arms/1177732360}.
\newblock URL \url{https://doi.org/10.1214/aoms/1177732360}.

\bibitem[Wood(2013)]{wood2013simple}
Simon~N Wood.
\newblock A simple test for random effects in regression models.
\newblock \emph{Biometrika}, 100\penalty0 (4):\penalty0 1005--1010, 2013.

\bibitem[Zhou et~al.(2022)Zhou, Heuvelink, Kono, Matsui, and Tanaka]{zhou2022}
Xinbin Zhou, Gerard~B.M. Heuvelink, Yusuke Kono, Tsutomu Matsui, and Takashi~S.T. Tanaka.
\newblock Using linear mixed-effects modeling to evaluate the impact of edaphic factors on spatial variation in winter wheat grain yield in japanese consolidated paddy fields.
\newblock \emph{European Journal of Agronomy}, 133:\penalty0 126447, 2022.
\newblock ISSN 1161-0301.

\end{thebibliography}
\end{appendices}

\end{document}